\documentclass[12pt, conference]{IEEEtran}

\usepackage{macro}
\usepackage[backend=biber]{biblatex}
\onecolumn
\title{Neural Polar Decoders for DNA Data Storage}

\author{
 \IEEEauthorblockN{Ziv Aharoni and Henry D. Pfister}\\
\IEEEauthorblockA{Department of Electrical and Computer Engineering\\
                   Duke University\\
                   Email: \{ziv.aharoni,henry.pfister\}@duke.edu}}
\begin{document}
\maketitle
\glsdisablehyper

\begin{abstract}
	Synchronization errors, such as insertions and deletions, present a fundamental challenge in DNA-based data storage systems, arising from both synthesis and sequencing noise. These channels are often modeled as \gls{ids} channels, for which designing maximum-likelihood decoders is computationally expensive. 
	In this work, we propose a data-driven approach based on \glspl{npd} to design low-complexity decoders for channels with synchronization errors. The proposed architecture enables decoding over \gls{ids} channels with reduced complexity $O(A N \log N)$, where $A$ is a tunable parameter independent of the channel. \Glspl{npd} require only sample access to the channel and can be trained without an explicit channel model. Additionally, \glspl{npd} provide \gls{mi} estimates that can be used to optimize input distributions and code design. 
	We demonstrate the effectiveness of \glspl{npd} on both synthetic deletion and \gls{ids} channels. For deletion channels, we show that \glspl{npd} achieve near-optimal decoding performance and accurate \gls{mi} estimation, with significantly lower complexity than trellis-based decoders. We also provide numerical estimates of the channel capacity for the deletion channel. We extend our evaluation to realistic DNA storage settings, including channels with multiple noisy reads and real-world Nanopore sequencing data. Our results show that \glspl{npd} match or surpass the performance of existing methods while using significantly fewer parameters than the state-of-the-art. These findings highlight the promise of \glspl{npd} for robust and efficient decoding in DNA data storage systems.
	\end{abstract}

\glsresetall
\blfootnote{This research was supported in part by NSF Grant Number 2212437.  Any opinions, findings, recommendations, and conclusions expressed in this material are those of the authors and do not necessarily reflect the views of the sponsor.}
\section{Introduction}
\par DNA-based data storage is an emerging paradigm that encodes digital data into synthetic DNA sequences \cite{bancroftLongtermStorageInformation2001,leproustSynthesisHighqualityLibraries2010,churchNextgenerationDigitalInformation2012,grassRobustChemicalPreservation2015,heckelCharacterizationDNAData2018,organickRandomAccessLargescale2018,meiserSyntheticDNAApplications2022}. It offers extremely high storage density and long-term durability, but also introduces unique noise characteristics stemming from synthesis, storage, and sequencing processes. Two major challenges in this domain are the presence of synchronization errors—particularly insertions and deletions—which are typically modeled by \gls{ids} channels, and the fact that strands are drawn randomly from a pool for reading. Each DNA strand may be sequenced multiple times, producing multiple noisy observations known as traces \cite{lenzCodingSetsDNA2020,lenzAchievingCapacityDNA2020,daveyReliableCommunicationChannels2001}. Moreover, DNA strands are relatively short \cite{leproustSynthesisHighqualityLibraries2010}, which necessitates appropriate coding schemes to ensure reliable data retrieval.

\par The DNA storage pipeline \cite{erlichDNAFountainEnables2017,heckelCharacterizationDNAData2018,heckelFundamentalLimitsDNA2017a,srinivasavaradhanTrellisBMACoded2024,bar-levScalableRobustDNAbased2025,sabarySurveyDecadeCoding2024} typically encodes digital data into multiple short sequences over a $4$-ary alphabet, which are synthesized and stored in an unordered pool. During retrieval, sequence selection is performed randomly, typically modeled as a Poisson sampling process. As a result, each stored strand may be sequenced multiple times, a few times, or not at allOnce a sequencing threshold is met (for example, when the average number of traces per stored sequence exceeds a value $c > 0$), the sequenced outputs are used to reconstruct the original data. This inherent variability in the sequencing process necessitates the use of both inner and outer codes: the inner code ensures reliable decoding from a limited number of traces, while the outer code corrects for strands that are not sampled. Addressing this variability has been a major focus in recent research \cite{erlichDNAFountainEnables2017,organickRandomAccessLargescale2018,srinivasavaradhanTrellisBMACoded2024,bar-levScalableRobustDNAbased2025}.

\par Towards designing efficient decoders for DNA data storage systems, we first consider the family of \gls{ids} channels, and deletion channels in particular. These channels were first studied in \cite{gallagerSequentialDecodingBinary1961,dobrushinShannonsTheoremsChannels1967}, and modern coding techniques were first applied to them in \cite{daveyReliableCommunicationChannels2001}. The difficulty stems from the exponential number of possible alignments between input and output sequences; consequently, many solutions in the literature were able to recover only from a fixed number of insertions or deletions \cite{varshamovCodesWhichCorrect1965a,daveyReliableCommunicationChannels2001,levenshteinEfficientReconstructionSequences2001,batu2004reconstructing}.

\par Recently, polar codes were applied to deletion channels \cite{talPolarCodesDeletion2021} and shown to achieve capacity when using a sequence of Markov input distributions. The main idea is to construct a trellis-based \gls{sc} decoder that computes the exact posteriors of the transmitted bits. The main drawback of this approach is its high decoding complexity: for a uniform input distribution, it is $O(N^4\log N)$, where $N$ is the block length. For a Markov input distribution with $S$ states, the complexity grows by an additional factor of $S^3$. This limits the practical usage of the approach for large block lengths. Nevertheless, even if the computational complexity were resolved, the method can only construct a code for a fixed Markov input distribution and does not provide a method to find the capacity-achieving one. In addition, the method requires a known channel model, which is unavailable in practice for DNA storage.

\par In recent work, a data-driven decoding approach based on \glspl{npd} \cite{aharoniDatadrivenNeuralPolar2024} was proposed for \glspl{fsc}. An \gls{npd} retains the recursive structure of a polar \gls{sc} decoder but replaces its elementary operations with \glspl{nn}. The parameters of the \gls{npd} are learned directly from input-output samples without requiring an explicit channel model. Moreover, the size of the \glspl{nn} associated with the \gls{npd} determines the decoding complexity $O(AN\log N)$, where $A$ quantifies the computational complexity of the realized \glspl{nn}. In \cite{aharoniCodeRateOptimization2024}, the \gls{npd} is used to estimate the \gls{mi} between the input and output sequences, which is then leveraged to optimize the input distribution and code design.

\par In this work, we present an \gls{npd}-based, data-driven approach to address DNA storage systems. The proposed solution first extends the \gls{npd} to channels with synchronization errors and multiple traces. The extension is verified on deletion channels and \gls{ids} channels, for which theoretical bounds are available in the literature. In particular, for deletion channels, we compare the decoding errors of the \gls{npd} with those of the trellis-based decoder from \cite{talPolarCodesDeletion2021}. We also leverage the \gls{npd} to optimize the input distribution and compare the optimized \gls{mi} with theoretical lower bounds \cite{castiglioneTrellisBasedLower2015} and upper bounds \cite{rahmatiUpperBoundsCapacity2015} on the deletion channel capacity. Finally, we extend our evaluation to \gls{ids} channels, where we compare the \gls{mi} attained by the \gls{npd} with the bounds reported in \cite{fertonaniBoundsCapacityChannels2011}.

\par After establishing the \gls{npd} on \gls{ids} channels, we proceed to evaluate the \gls{npd} in realistic DNA storage settings. We first consider synthetic DNA storage channels, modeled as \gls{ids} channels over a $4$-ary alphabet with $K \sim \operatorname{Poiss}(\lambda)$ independent traces. For this setting, we demonstrate how to use the \gls{npd} to jointly design the inner and outer codes to mitigate the variability in the number of traces obtained from the sequencing process. We then extend the evaluation to real-world DNA storage data, using the Nanopore dataset from \cite{srinivasavaradhanTrellisBMACoded2024}. In this case, we show that the \gls{npd} achieves performance competitive with state-of-the-art methods \cite{bar-levScalableRobustDNAbased2025}, while using significantly fewer parameters.

\par The paper is organized as follows. In Section \ref{sec:background}, we provide notations and preliminaries. In Section \ref{sec:npd}, we present the \gls{npd} architecture and its training process. In Section \ref{sec:npd_extend}, we extend the \gls{npd} to channels with synchronization errors and multiple traces. In Section \ref{sec:experiments_synthetic}, we present the experimental results on deletion and \gls{ids} channels. In Section \ref{sec:experiments_dna}, we present the experimental results on DNA storage channels. Finally, we conclude the paper in Section \ref{sec:conclusion}.

\section{Background}\label{sec:background}
\par In this section, we present the notation and preliminaries used throughout the paper. We begin with the notation and definitions of \gls{ids} channels, followed by the definitions of polar codes and \gls{sc} decoders.

\subsection{Notation}
\par We denote by $(\Omega,\mathcal{F},\mathbb{P})$ the underlying probability space on which all \glspl{rv} are defined. Sets are denoted by calligraphic letters, e.g., $\mathcal{X}$. 
We use the notation $\mathbf{X}$ to denote the random vector $(X_1, X_2, \dots, X_N)$ and $\mathbf{x}$ to denote its realization, where $N$ will be clear from context. The term $P_{\mathbf{X}}$ denotes the distribution of $\mathbf{X}$.
For two sets $A \subseteq B$, $B \setminus A$ denotes the set difference operation.
We denote by $[N]$ the set of integers $\{1,\dots,N\}$. For $\mathcal{D} \subseteq [N]$, let $\mathcal{D}^c = [N] \setminus \mathcal{D}$ denote its complement. 
Let $\mathbf{x}$ be a length-$N$ binary sequence. The support of $\mathbf{x}$ is defined as $\operatorname{supp}(\mathbf{x}) = \{ i \mid x_i = 1, \; i \in [N] \}$.
Given an index set $\mathcal{D} \subseteq [N]$, the notation $\mathbf{X}_{\mathcal{D}}$ represents the subsequence of $\mathbf{X}$ consisting of the elements indexed by $\mathcal{D}$, i.e., $\mathbf{X}_{\mathcal{D}} = (X_i \mid i \in \mathcal{D})$, with the elements of $\mathcal{D}$ in increasing order. 
The term $A \otimes B$ denotes the Kronecker product of $A$ and $B$ when $A, B$ are matrices, and the product distribution when $A, B$ are distributions. 
The notation $A^{\otimes N} := A \otimes A \otimes \dots \otimes A$ denotes applying the $\otimes$ operator $N$ times.  
The notation $\mathbf{D} \sim \mathsf{Ber}(p)^{\otimes N}$ indicates that $\mathbf{D}$ is a length-$N$ \gls{iid} Bernoulli random vector with success rate $p \in (0,1)$.
Let $\mathbf{X} \in \mathcal{X}^N$ and $\mathbf{Y} \in \Sigma^M$ be the inputs and outputs of a channel, respectively, where the input alphabet $\mathcal{X}$ and the output alphabet $\Sigma$ are finite. Their joint distribution factors as $P_{\mathbf{X}} \otimes W_{\mathbf{Y}|\mathbf{X}}$, and is explicitly given by
$P_{\mathbf{X},\mathbf{Y}}(\mathbf{x},\mathbf{y}) =  P_{\mathbf{X}}(\mathbf{x}) W(\mathbf{y} | \mathbf{x})$,
emphasizing that there is no feedback between the encoder and decoder, and that $\mathbf{x}$ and $\mathbf{y}$ are generally sequences of different lengths.
The binary entropy of $X$ is denoted by $\f{\sH_2}{X}$, the \gls{ce} between two distributions $P$ and $Q$ is denoted by $\ce{P}{Q}$, and the \gls{mi} between two \glspl{rv} $X$ and $Y$ is denoted by $\f{\sI}{X;Y}$.

\subsection{Channel Definitions}
Next, we define the \gls{ids} channel, following the approach previously used in \cite{srinivasavaradhanTrellisBMACoded2024,daveyReliableCommunicationChannels2001}.

\begin{definition}[Insertion-deletion-substitution channel]\label{def:ids}
An \gls{ids} channel $\operatorname{IDS}(i,d,s)$ is defined by parameters $i,d,s \in [0,1]$, which designate the insertion, deletion, and substitution probabilities, respectively, with $i+d+s \leq 1$. Given $\mathbf{x} \in \Sigma^N$ with $|\Sigma| < \infty$, the output of the \gls{ids} channel is $\mathbf{y} \in \Sigma^\ast$, where $\Sigma^\ast = \cup_{n \in \mathbb{N}} \Sigma^n$. The output is computed sequentially via an auxiliary sequence of independent \glspl{rv} $Z^N$, where each $Z_n$ takes values in $\{\mathsf{ins}, \mathsf{del}, \mathsf{sub}, \mathsf{clean}\}$ with probabilities $i$, $d$, $s$, and $1-i-d-s$, respectively. The output $\mathbf{y}$ is generated as follows, starting with input pointer $n=1$ and output pointer $m=1$:
\begin{itemize}
    \item If $n > N$, the process stops.
    \item If $Z_n = \mathsf{ins}$, then $y_m \sim \mathsf{Unif}(\Sigma)$ and $m \leftarrow m+1$.
    \item If $Z_n = \mathsf{del}$, then $n \leftarrow n+1$.
    \item If $Z_n = \mathsf{sub}$, then $y_m \sim \mathsf{Unif}(\Sigma \setminus \{x_n\})$ and $m,n \leftarrow m+1,n+1$.
    \item If $Z_n = \mathsf{clean}$, then $y_m = x_n$ and $m,n \leftarrow m+1,n+1$.
\end{itemize}
\end{definition}

\par From this definition, it is straightforward to define the deletion channel by $\operatorname{Del}(d) \triangleq \operatorname{IDS}(0,d,0)$. Based on Definition~\ref{def:ids}, we now define the DNA channel model.

\begin{definition}[DNA channel model]
The DNA channel model is specified by the tuple $(i,d,s,\lambda)$, where $i+d+s \leq 1$ and $\lambda > 0$. Let $W$ denote the transition kernel of $\operatorname{IDS}(i,d,s)$. Given $\mathbf{x} \in \Sigma^N$, the output of the DNA channel is generated as follows:
\begin{itemize}
    \item The number of traces $K$ is sampled according to $K \sim \operatorname{Poiss}(\lambda)$.
    \item Each trace is independently sampled as $\mathbf{y}_k \sim W(\cdot|\mathbf{x})$ for $k \in [K]$.
\end{itemize}
The output of the DNA channel is denoted by $\mathbf{y}^{1:K} = (\mathbf{y}_1, \dots, \mathbf{y}_K)$.
\end{definition}

\subsection{Polar Codes}
\par Let $G_N = B_N F^{\otimes n}$ denote Ar\i{}kan's polar transform, where $G_N$ is the generator matrix for block length $N=2^n$ with $n\in\mathbb{N}$. The matrix $F$ is Ar\i{}kan's $2\times2$ kernel defined by
\[
    F = \begin{bmatrix}
    1 & 0 \\
    1 & 1
    \end{bmatrix}.
\]
The matrix $B_N$ is the permutation matrix associated with the bit-reversal permutation. It is defined recursively by $B_N = R_N (I_2 \otimes B_{N/2})$, starting from $B_2 = I_2$.
Here, $I_N$ denotes the identity matrix of size $N$, and $R_N$ denotes a permutation matrix called the reverse-shuffle permutation \cite{arikanChannelPolarizationMethod2009}.
Given $\mathbf{x} \in \mathbb{F}_2^N$, the polar transform is defined by
\begin{equation}
    \mathbf{u} = \mathbf{x} G_N.
\end{equation}
The synthetic channels of the polar code are defined by
\begin{equation}\label{eqn:synthetic_channels}
W_{i,N}(u_i|u^{i-1},y^N) = P_{U_i|U^{i-1},Y^N}(u_i|u^{i-1},y^N),
\end{equation}
for $i \in [N]$, $\mathbf{u} \in \mathcal{X}^N$, and $\mathbf{y} \in \mathcal{Y}^N$. 

\par A \gls{sc} polar decoder is composed of four elementary functions that map the channel outputs $\mathbf{y}$ into the estimated information vector $\hat{\mathbf{u}}$ via the computation of the synthetic channels \cite{arikanChannelPolarizationMethod2009,aharoniDatadrivenNeuralPolar2024}. Thus, a \gls{sc} polar decoder may be defined as follows.

\begin{definition}[Successive cancellation polar decoder]\label{def:sc_polar_decoder}
Let $\mathbf{x} \in \mathcal{X}^N$, $\mathbf{y} \in \mathcal{Y}^N$, and $\mathcal{E} \subset \mathbb{R}^d$, where $d \in \mathbb{N}$. A \gls{sc} polar decoder is composed of the following functions:
\begin{itemize}
    \item The embedding function $E: \mathcal{Y} \to \mathcal{E}$,
    \item The check-node function $F: \mathcal{E} \times \mathcal{E} \to \mathcal{E}$,
    \item The bit-node function $G: \mathcal{E} \times \mathcal{E} \times \mathcal{X} \to \mathcal{E}$,
    \item The embedding-to-LLR function $H: \mathcal{E} \to \mathbb{R}$.
\end{itemize}
\end{definition}

\par Definition~\ref{def:sc_polar_decoder} lists the ingredients needed by a \gls{sc} decoder to decode an output sequence $\mathbf{y} \in \mathcal{X}^{N \times 1}$ into an estimate $\hat{\mathbf{u}} \in \mathcal{X}^{N \times 1}$. The decoding computation starts by mapping each $y \in \mathbf{y}$ independently into $e = E(y)$, where $e \in \mathbb{R}^{1 \times d}$, to form $\mathbf{e} \in \mathbb{R}^{N \times d}$. The decoding is then completed by using $\mathbf{e}$ and the functions $F, G, H$ to compute $\hat{\mathbf{u}}$ via the recursion of the \gls{sc} decoder, as shown in \cite{arikanChannelPolarizationMethod2009}. 

For example, setting $\mathcal{E} = \mathbb{R}$ and defining $E, F, G, H$ as
\begin{align}\label{eqn:scd_memoryless}
    E(y) &= \log\frac{W(y|1)}{W(y|0)} + \log\frac{P_X(1)}{P_X(0)}, \nonumber\\
    F(e_1, e_2) &= -2\tanh^{-1}\left(\tanh\left(\frac{e_1}{2}\right)\tanh\left(\frac{e_2}{2}\right)\right), \nonumber\\
    G(e_1, e_2, u) &= e_2 + (-1)^u e_1, \nonumber\\
    H(e_1) &= e_1,
\end{align}
for $e_1, e_2 \in \mathbb{R}$ and $u \in \mathcal{X}$, defines an \gls{sc} decoder for binary-input memoryless channels.

For \glspl{fsc} with state space $\mathcal{S}$, the trellis-based decoder from \cite{wangConstructionPolarCodes2015} is defined by setting $\mathcal{E} = [0,1]^{|\mathcal{X}||\mathcal{S}|^2}$ and defining
\begin{align}\label{eqn:scd_fscs}
    [E(y)]_{x,s_0,s_1} = P_{X,Y,S'|S}(x,y,s_1|s_0).
\end{align}
Here, the coordinates $(x, s_0, s_1)$ specify an entry of the vector $e$, and $y$ is the input to the embedding function $E$. The embedding function outputs a vector—commonly referred to as a trellis—that captures the joint distribution of $(X, Y)$ and the channel states $(S, S')$, representing the state before and after transmission, respectively. The corresponding functions $F, G, H$ are provided in \cite{wangConstructionPolarCodes2015}.

\section{Neural Polar Decoders}\label{sec:npd}
\par An \gls{npd} is also an instance of an \gls{sc} polar decoder, as defined in Definition~\ref{def:sc_polar_decoder}. Setting $\mathcal{E} = \mathbb{R}^d$, with $d \in \mathbb{N}$, and realizing the functions $E, F, G, H$ using \glspl{nn} yields an \gls{npd}. Let $\mathcal{G}_\mathsf{NN}(d_i, k, d_o)$ denote the family of shallow \glspl{nn} with $d_i$ inputs, $k$ hidden units, $d_o$ outputs, and ReLU activations. For additional details, the reader is referred to \cite{aharoniDatadrivenNeuralPolar2024}.

\begin{definition}[Neural polar decoder]\label{def:npd}
An \gls{npd} is an \gls{sc} polar decoder, as defined in Definition~\ref{def:sc_polar_decoder}, with the functions $E, F, G, H$ defined by:
\begin{itemize}
    \item The embedding \gls{nn} $E_{\theta_E} \in \mathcal{G}_\mathsf{NN}(1, h, d)$,
    \item The check-node \gls{nn} $F_{\theta_F} \in \mathcal{G}_\mathsf{NN}(2d, h, d)$,
    \item The bit-node \gls{nn} $G_{\theta_G} \in \mathcal{G}_\mathsf{NN}(2d+1, h, d)$,
    \item The embedding-to-LLR \gls{nn} $H_{\theta_H} \in \mathcal{G}_\mathsf{NN}(d, h, 1)$,
\end{itemize}
where $h \in \mathbb{N}$ is the number of hidden units. Let $\theta = \left\{\theta_E, \theta_F, \theta_G, \theta_H\right\}$ be the parameters of the \gls{npd}, and $\Theta$ the corresponding parameter space.
\end{definition}

\subsection{Learning the NPD's Parameters}
\par For any $\theta \in \Theta$, the \gls{npd} computes an estimate of the posterior distribution of the synthetic channels $W_{i,N}$, as defined in \eqref{eqn:synthetic_channels}, via the recursion of the \gls{sc} decoder. Let this estimate be denoted by $W^\theta_{i,N}$. The process of determining the \gls{npd}'s parameters is a \gls{sgd} optimization procedure, in which examples of input-output pairs are used to optimize the parameters $\theta$. 

The optimization process minimizes the sum of cross-entropy terms:
\begin{equation}\label{eqn:npd_loss}
    \min_{\theta \in \Theta} \frac{1}{N} \sum_{i=1}^N \ce{W_{i,N}}{W_{i,N}^\theta},
\end{equation}
which achieves its minimum value of $\frac{1}{N}\sum_{i=1}^N \f{\sH_2}{W_{i,N}}$ if and only if $W_{i,N} = W^\theta_{i,N}$ almost surely.

The \gls{ce} terms are approximated via the empirical negative log-loss function. Specifically, given a dataset $\mathcal{S}_M = \{ (\mathbf{x}_i, \mathbf{y}_i) \}_{i=1}^M \stackrel{iid}{\sim} P_{\mathbf{X}} \otimes W_{\mathbf{Y}|\mathbf{X}}$, the loss in \eqref{eqn:npd_loss} is approximated by
\begin{equation}
    \min_{\theta \in \Theta} \frac{1}{M} \sum_{(\mathbf{x}, \mathbf{y}) \in \mathcal{S}_M} \mathbf{l}(\mathbf{x}, \mathbf{y}; \theta),
\end{equation}
where
\[
\mathbf{l}(\mathbf{x}, \mathbf{y}; \theta) = \frac{1}{N} \sum_{i=1}^N -\log W_{i,N}^\theta \left( u_{i} \mid u_{1}^{i-1}, \mathbf{y} \right),
\]
with $\mathbf{u} = \mathbf{x} G_N$ and $u_{1}^{i-1} = (u_1, \dots, u_{i-1})$.
The full optimization procedure is outlined in \cite{aharoniDatadrivenNeuralPolar2024}.

\subsection{Estimation of Mutual Information via NPDs}\label{sec:npd_mi_est}
\par Let $P_\mathbf{X}^\psi$ denote the input distribution parameterized by $\psi$, and let $\f{\sI_\psi}{\mathbf{X};\mathbf{Y}}$ denote the \gls{mi} between $\mathbf{X}$ and $\mathbf{Y}$, where the subscript $\psi$ highlights the dependence of the mutual information on $\psi$. Let $\mathcal{S}_M^\psi = \{ (\mathbf{x}_i, \mathbf{y}_i) \}_{i=1}^M \stackrel{iid}{\sim} P_\mathbf{X}^\psi \otimes W_{\mathbf{Y}|\mathbf{X}}$. The mutual information is estimated by applying the \gls{npd} twice to estimate the two conditional entropies appearing on the right-hand side of
\begin{equation}\label{eqn:mi_hy}
    \f{\sI_\psi}{\mathbf{X};\mathbf{Y}} = \f{\sH_\psi}{\mathbf{U}} - \f{\sH_\psi}{\mathbf{U}|\mathbf{Y}},
\end{equation}
where the equality holds because $\mathbf{u} = \mathbf{x}G_N$ is a bijective mapping. From \eqref{eqn:mi_hy}, it follows that the mutual information can be estimated by applying the \gls{npd} twice: once on the original channel $W_{\mathbf{Y}|\mathbf{X}}$, and once on a modified channel $\widetilde{W}_{\mathbf{Y}|\mathbf{X}}$ where $\mathbf{X} \indep \mathbf{Y}$, e.g. $\mathbf{Y}=\mathbf{0}$. The first entropy $\f{\sH_\psi}{\mathbf{U}|\mathbf{Y}}$ is estimated using an \gls{npd} with parameters $\theta_W = \{ \theta_E, \theta_F, \theta_G, \theta_H \}$, while the second entropy $\f{\sH_\psi}{\mathbf{U}}$ is estimated using an \gls{npd} with parameters $\theta_{\widetilde{W}} = \{ \theta_{\tilde{E}}, \theta_F, \theta_G, \theta_H \}$, where the networks corresponding to $F$, $G$, and $H$ are shared between the two decoders. The training objective for estimating both entropies is given by
\begin{equation}\label{eqn:mi_est_obj}
    \min_{\theta_W,\theta_{\widetilde{W}}} \frac{1}{M}\sum_{(\mathbf{x},\mathbf{y})\in\mathcal{S}_M^\psi} \mathbf{l}(\mathbf{x}, \mathbf{y}; \theta_W) + \frac{1}{M}\sum_{(\mathbf{x},\mathbf{y})\in\mathcal{S}_M^\psi} \mathbf{l}(\mathbf{x}, \mathbf{0}; \theta_{\widetilde{W}}),
\end{equation}
where the overall parameter set is $\theta = \{ \theta_E, \theta_{\tilde{E}}, \theta_F, \theta_G, \theta_H \}$. For given parameters $\theta = \{ \theta_W, \theta_{\widetilde{W}} \}$ and input distribution parameters $\psi$, the estimated mutual information is computed as
\begin{equation}\label{eqn:mi_est}
    \f{\hat{\sI}_{\psi}^{\theta}}{\mathbf{X};\mathbf{Y}} = \frac{1}{M} \sum_{(\mathbf{x},\mathbf{y})\in\mathcal{S}_M^\psi} \Big( \mathbf{l}(\mathbf{x}, \mathbf{0}; \theta_{\widetilde{W}}) - \mathbf{l}(\mathbf{x}, \mathbf{y}; \theta_W) \Big),
\end{equation}
where the superscript $\theta$ indicates the dependence of the estimate on the NPD parameters, and the subscript $\psi$ highlights the dependence on the input distribution parameters. The difference $\mathbf{l}(\mathbf{x}, \mathbf{0}; \theta_{\widetilde{W}}) - \mathbf{l}(\mathbf{x}, \mathbf{y}; \theta_W)$ serves as a proxy of the average information density of the synthetic channels $\frac{1}{N}\sum_{i=1}^N\log\frac{P_{U_i|U^{i-1},Y^N}}{P_{U_i|U^{i-1}}}$, whose average over $\mathcal{S}_M^\psi$ approximates $ \f{\sI_\psi}{\mathbf{X};\mathbf{Y}}$.

\subsection{Optimizing the Input Distribution}\label{sec:npd_mi_opt}
\par The input distribution optimization \cite{aharoniCodeRateOptimization2024} is performed via an alternating maximization scheme, in which the parameters of the \gls{npd}, $\theta$, and the parameters of the input distribution, $\psi$, are optimized interchangeably. The algorithm begins with random initialization of $\theta^{(0)}$ and $\psi^{(0)}$, and starts with a warm-up phase where $\psi^{(0)}$ is fixed. A sample $\mathcal{S}_M^{\psi^{(0)}} \stackrel{iid}{\sim} P_{\mathbf{X}}^{\psi^{(0)}} \otimes W_{\mathbf{Y}|\mathbf{X}}$ is drawn, and the objective in \eqref{eqn:mi_est_obj} is optimized to yield $\theta^{(1)}$, which provides an estimate of the \gls{mi} induced by $\psi^{(0)}$, $\f{\hat{\sI}_{\psi^{(0)}}^{\theta^{(1)}}}{\mathbf{X};\mathbf{Y}}$.

\par Next, $\theta^{(1)}$ is fixed, and the gradient with respect to the input distribution parameters is computed as
\begin{equation}\label{eqn:input_gradient}
    \nabla_\psi \f{\hat{\sI}_{\psi}^{\theta}}{\mathbf{X};\mathbf{Y}} = \frac{1}{M} \sum_{(\mathbf{x},\mathbf{y}) \in \mathcal{S}_M^\psi} \nabla_\psi \log P_{\mathbf{X}}^\psi(\mathbf{x}) \Big( \mathbf{l}(\mathbf{x},\mathbf{0};\theta_{\widetilde{W}}) - \mathbf{l}(\mathbf{x},\mathbf{y};\theta_W) \Big),
\end{equation}
which was shown in \cite{aharoniCodeRateOptimization2024} to be a consistent estimator of $\nabla_\psi \sI_\psi(\mathbf{X};\mathbf{Y})$. Equation \eqref{eqn:input_gradient} can be interpreted as a weighted maximum likelihood update of the input distribution, where sequences with higher information density receive higher weights.

\par To improve empirical convergence, in practice, we normalize the gradient using the following objective:
\begin{equation}\label{eqn:input_gradient_practical}
    \nabla_\psi \f{\hat{\sI}_{\psi}^{\theta}}{\mathbf{X};\mathbf{Y}} = \frac{1}{M} \sum_{(\mathbf{x},\mathbf{y}) \in \mathcal{S}_M^\psi} \nabla_\psi \log P_{\mathbf{X}}^\psi(\mathbf{x}) \left( \frac{r(\mathbf{x},\mathbf{y};\theta) - \f{\hat{\sI}_{\psi}^{\theta}}{\mathbf{X};\mathbf{Y}}}{\sqrt{\frac{1}{M}\sum_{(\mathbf{x}',\mathbf{y}') \in \mathcal{S}_M^\psi} \left( r(\mathbf{x}',\mathbf{y}';\theta) - \f{\hat{\sI}_{\psi}^{\theta}}{\mathbf{X};\mathbf{Y}} \right)^2}} \right),
\end{equation}
where $r(\mathbf{x},\mathbf{y};\theta) = \mathbf{l}(\mathbf{x},\mathbf{0};\theta_{\widetilde{W}}) - \mathbf{l}(\mathbf{x},\mathbf{y};\theta_W)$. The difference between \eqref{eqn:input_gradient} and \eqref{eqn:input_gradient_practical} lies in the normalization of the estimates of the information density. This normalization technique is commonly used in reinforcement learning algorithms \cite{mnihAsynchronousMethodsDeep2016}, and it has been observed that normalizing the rewards to have zero mean and unit standard deviation yields better empirical performance.

\par After applying the gradient update from \eqref{eqn:input_gradient_practical}, we obtain $\psi^{(1)}$. The procedure then continues iteratively: at iteration $k$, the algorithm draws a sample $\mathcal{S}_M^{\psi^{(k-1)}}$, optimizes the objective in \eqref{eqn:mi_est_obj} to obtain $\theta^{(k)}$, and then computes the gradient of the input distribution to update and obtain $\psi^{(k)}$. This process continues for a predetermined number of iterations or until the \gls{mi} estimates converge and stop improving.

\section{Neural Polar Decoder for Channels with Synchronization Errors}\label{sec:npd_extend}
\par This section addresses the generalization of \glspl{npd} to \gls{ids} channels. It begins by discussing the limitations of current \glspl{npd} when applied to channels with synchronization errors. It then focuses on the modification of the embedding \gls{nn} $E_\theta$ to address these challenges. Finally, the section presents a method for computing embeddings when multiple traces are available.

\subsection{Extending the Embedding NN}
\par Generally, the embedding $e_i \in \mathbb{R}^d$ serves as the belief about the transmitted symbol $x_i$. For memoryless channels and \glspl{fsc}, the belief of $x_i$ may be computed based only on $y_i$, as given in \eqref{eqn:scd_memoryless} and \eqref{eqn:scd_fscs}, respectively. As shown in \cite{aharoniDatadrivenNeuralPolar2024}, the belief may be approximated by the embedding \gls{nn}. However, this approach cannot be applied directly to channels with synchronization errors, for the obvious reason that there is no longer a one-to-one correspondence between each $x_i$ and a single observation.

\par To address this issue, we draw motivation from the trellis-based \gls{sc} decoder for deletion channels proposed in \cite{talPolarCodesDeletion2021}. In that work, the trellis, which captures the belief about $\mathbf{x}$ given $\mathbf{y}$, is constructed by considering multiple observed outputs from $\mathbf{y}$, rather than a single output per input symbol as done in \cite{arikanChannelPolarizationMethod2009,wangConstructionPolarCodes2015}. Thus, instead of computing $\mathbf{e} \in \mathbb{R}^{N \times d}$ by applying the embedding \gls{nn} independently as $e = E_\theta(y)$ for each $y \in \mathbf{y}$, we compute $\mathbf{e} = \mathbf{E}_\theta(\mathbf{y})$, where $\mathbf{E}_\theta$ is a \gls{nn} that maps the entire $\mathbf{y}$ vector into $\mathbf{e}$.


\begin{figure}
	\centering
	\resizebox{7cm}{!}{
	\begin{tikzpicture}
        \foreach \x in {0.5, 2.5, 4.5, 6.5} {
            \fill[yellow!50] (\x, -3.5) rectangle ++(1, 7.5);
        }
        \node[] at (1, 4.5) {$x_1$};
        \node[] at (3, 4.5) {$x_2$};
        \node[] at (5, 4.5) {$x_3$};
        \node[] at (7, 4.5) {$x_4$};
        \node[blue] at (-1, 2) {$y_1 = 1$};
        \node[red] at (-1, 0) {$y_2 = 0$};
        \node[blue] at (-1, -2) {$y_3 = 1$};
    
        \foreach \y in {3, 1, -1, -3} {
            \foreach \x in {0, 2, 4, 6, 8} {
                \node[circle, fill=black, inner sep=1pt] at (\x, \y) {};
            }
        }
        \draw (0, 3) circle (0.15cm);
        \draw (8, -3) circle (0.15cm);
        \draw[->, red, thick] (0, 3) to[out=45, in=135]node[midway, above] {$\scriptstyle\delta / 2$}(2, 3);
        \draw[->, blue, thick] (0, 3) to[out=-45, in=-135]node[midway, above] {$\scriptstyle\delta / 2$}(2, 3);
        \draw[->, blue, thick] (0, 3) tonode[midway, above] {$\scriptstyle\bar{\delta} / 2$}(2, 1);
        \draw[->, blue, thick] (2, 3) tonode[midway, above] {$\scriptstyle\bar{\delta} / 2$}(4, 1);
        
        \draw[->, red, thick] (2, 1) to[out=45, in=135]node[midway, above] {$\scriptstyle\delta / 2$}(4, 1);
        \draw[->, blue, thick] (2, 1) to[out=-45, in=-135]node[midway, above] {$\scriptstyle\delta / 2$}(4, 1);
        \draw[->, red, thick] (2, 1) tonode[midway, above] {$\scriptstyle\bar{\delta} / 2$}(4, -1);
        \draw[->, red, thick] (4, 1) tonode[midway, above] {$\scriptstyle\bar{\delta} / 2$}(6, -1);

        \draw[->, red, thick] (4, -1) to[out=45, in=135]node[midway, above] {$\scriptstyle\delta / 2$}(6, -1);
        \draw[->, blue, thick] (4, -1) to[out=-45, in=-135]node[midway, above] {$\scriptstyle\delta / 2$}(6, -1);
        \draw[->, blue, thick] (4, -1) tonode[midway, above] {$\scriptstyle\bar{\delta} / 2$}(6, -3);
        \draw[->, blue, thick] (6, -1) tonode[midway, above] {$\scriptstyle\bar{\delta} / 2$}(8, -3);
        
        \draw[->, red, thick] (6, -3) to[out=45, in=135]node[midway, above] {$\scriptstyle\delta / 2$}(8, -3);
        \draw[->, blue, thick] (6, -3) to[out=-45, in=-135]node[midway, above] {$\scriptstyle\delta / 2$}(8, -3);
    
    
    
        \node at (1, -4) {$e_1$};
        \node at (3, -4) {$e_2$};
        \node at (5, -4) {$e_3$};
        \node at (7, -4) {$e_4$};
    
    \end{tikzpicture}}
	\caption{The trellis in the \gls{sc} decoder for deletion channels \cite{talPolarCodesDeletion2021} with deletion rate $d$.}
	\label{fig:trellis-del}
\end{figure}
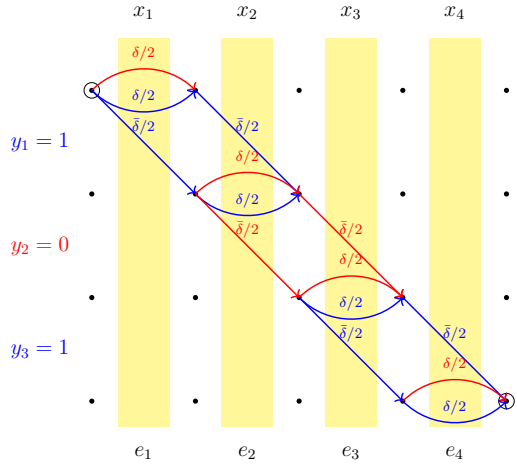

\subsection{Embedding NN Architecture}
\par The goal now is to describe two architectures for the embedding \gls{nn}. The first is a \gls{cnn}-based embedding, which requires fewer parameters and is primarily used for channels with only deletions. The second architecture employs an attention mechanism and is primarily used for \gls{ids} channels, as it better handles cases where the output length is either larger or smaller than $N$.

\par Both architectures are inspired by the trellis-based \gls{sc} decoder for deletion channels \cite{talPolarCodesDeletion2021}, illustrated in Figure~\ref{fig:trellis-del}. In the trellis, blue and red edges correspond to $x_i=1$ and $x_i=0$, respectively. Each complete path from the top-left to the bottom-right node represents a possible input sequence $\mathbf{x}$, and the product of the edge weights equals $P_{\mathbf{X}}(\mathbf{x}) W_{\mathbf{Y}|\mathbf{X}}(\mathbf{y}|\mathbf{x})$. Missing edges have weight zero. The edge weights within the $i$-th yellow stripe represent the belief for $x_i$. Thus, the trellis captures the joint distribution of $\mathbf{X}$ and $\mathbf{Y}$, providing beliefs for each transmitted bit $x_i$ that are fed into a polar \gls{sc} decoder tailored to this structure.

\par The trellis construction effectively mirrors the operation performed by the embedding function $E$. Therefore, in designing the embedding \gls{nn}, we use the trellis structure as guidance to capture the sufficient statistics required by the \gls{sc} decoder. Our main observations are twofold: first, the belief of the $i$-th bit depends explicitly on the index $i$, implying that the embedding function should encode positional information; second, the trellis edges depend on multiple output symbols. Thus, we design architectures that use positional embeddings and operate on the entire sequence $\mathbf{y}$.

\par The embedding computation proceeds as follows. The channel outputs $\mathbf{y} \in \Sigma^L$, with $L$ varying across different sequences, are first padded to a fixed-length sequence of size $L_\text{max} \ge N$ using an erasure symbol $``?"$. The padded sequence $\mathbf{\tilde{y}}$ is then embedded into a $d$-dimensional space by mapping each symbol in $\Sigma \cup \{?\}$ to a learned vector in $\mathbb{R}^d$. We denote this operation by $e = \operatorname{Embedding}(y)$ for $y \in \Sigma \cup \{?\}$. Given a matrix of embedded values $\mathbf{\tilde{e}} \in \mathbb{R}^{L_\text{max} \times d}$, positional encoding is incorporated by adding a learned positional embedding matrix $\mathbf{p} \in \mathbb{R}^{L_\text{max} \times d}$, following \cite{gehringConvolutionalSequenceSequence2017}. The resulting positionally-encoded embeddings $\mathbf{\tilde{e}}_\mathsf{pos}$ are then processed by a \gls{nn} to yield the final embeddings $\mathbf{e}$ used by the \gls{npd}.

\par The \gls{cnn}-based embedding is restricted to $L_\text{max} = N$, and is therefore mainly used for deletion channels. The architecture is defined as follows.
\begin{definition}[CNN Embedding]
Let $\mathbf{y} \in \Sigma^L$ be the channel outputs. The embedding \gls{nn} is defined by the following equations:
\begin{align*}
    \mathbf{\tilde{y}} &= (\mathbf{y},\mathbf{?}^{N-L}), && \text{(pad with erasure symbols)}\\
    \mathbf{\tilde{e}} &= \operatorname{Embedding}(\mathbf{\tilde{y}}), && \text{(embed channel outputs)}\\
    \mathbf{\tilde{e}}_\mathsf{pos} &= \mathbf{\tilde{e}} + \mathbf{p}, && \text{(add positional encoding)}\\
    \mathbf{e} &= \mathsf{CNN}(\mathbf{\tilde{e}}_\mathsf{pos}), && \text{(apply Convolutional NN)}
\end{align*}
where $(\mathbf{y},\mathbf{?}^{N-L})$ denotes the concatenation of erasure symbols to pad $\mathbf{y}$ to length $N$.
\end{definition}

\par For a general \gls{ids} channel, it is not guaranteed that $L \le N$, and therefore $L_\text{max}$ is chosen slightly larger than $N$, typically $10\%$ longer than the expected output length. However, the input to the \gls{npd} must be a matrix $\mathbf{e} \in \mathbb{R}^{N \times d}$ containing $N$ embedding vectors, representing the beliefs for the input symbols $\mathbf{x}$. Thus, it is necessary to transform the $L_\text{max}$ output symbols into $N$ embedding vectors, which is not possible with the \gls{cnn}-based embedding. To address this, we use an attention mechanism following \cite{vaswaniAttentionAllYou2017}.

\begin{definition}[Attention Embedding]\label{def:att_emb}
	Let $\mathbf{y} \in \Sigma^L$ be the channel outputs, and let $\mathbf{p}_\mathsf{in}$ denote the first $N$ rows of $\mathbf{p}$. The attention embedding is defined by the following equations:
	\begin{align*}
		\mathbf{\tilde{y}} &= (\mathbf{y},\mathbf{?}^{L_\text{max}-L}), && \text{(pad with erasure symbols)}\\
		\mathbf{\tilde{e}} &= \operatorname{Embedding}(\mathbf{\tilde{y}}), && \text{(embed channel outputs)}\\
		\mathbf{\tilde{e}}_\mathsf{pos} &= \mathbf{\tilde{e}} + \mathbf{p}, && \text{(add positional encoding)}\\
		\mathbf{e} &= \operatorname{AttentionNN}(\mathbf{p}_\mathsf{in},\mathbf{\tilde{e}}_\mathsf{pos},\mathbf{\tilde{e}}_\mathsf{pos}), && \text{(apply Attention NN)}
	\end{align*}
	where $\operatorname{AttentionNN}(q,k,v)$ denotes a two-layer attention-based \gls{nn} that produces embeddings in the required dimension for the decoder. 
\end{definition}
The attention mechanism performs the mapping $\operatorname{Softmax}\left( q k^T /\sqrt{d} \right) v$. 
For our choice of $q, k, v$ as in Definition~\ref{def:att_emb}, the outer product $q k^T \in \mathbb{R}^{N\times L_\text{max}}$, which, after computing the softmax and multiplying with the values matrix, results in $\mathbf{e} \in \mathbb{R}^{N\times d}$, as desired. More details on the implementation of both architectures are given in Appendix~\ref{app:implement}.

\subsection{Designing Embedding NN for Multiple Traces}
\par Due to the difficulties in synthesizing and sequencing long DNA molecules, DNA storage systems typically partition a file into many shorter strands \cite{leproustSynthesisHighqualityLibraries2010}. However, the Poisson nature of the sampling process means that, during sequencing, some strands may be sampled multiple times, while others may be missed entirely. Therefore, this section presents how to accommodate the case where the decoder observes independent versions of the channel output $\mathbf{y}^1, \dotsc, \mathbf{y}^K$.

\par Recall that $\mathbf{e} = H_\theta(\mathbf{y})$ defines the mapping of a single strand into the embedding used by the \gls{npd}. Under the assumption that the traces per strand are independent, we may view $\mathbf{y}^{1:K}$ as a repetition code over the sequence space $\Sigma^N$. In the memoryless case, such codes are decoded by summing the log-likelihood ratios across repetitions. Motivated by this analogy, we propose a similar aggregation mechanism for the \gls{npd}.

\par Specifically, to avoid introducing additional parameters into the embedding network, we define the embedding for multiple traces $\mathbf{y}^{1:K}$ as:
\begin{equation}\label{eqn:multiple_emb}
\mathbf{e} = \frac{1}{\sqrt{K}} \sum_{i=1}^K H_\theta(\mathbf{y}^i).
\end{equation}
The normalization by $\sqrt{K}$ aims to maintain a consistent second moment of the resulting embedding, independent of the number of traces $K$.

\par We note that this aggregation rule is clearly heuristic and is proposed as a practical design choice. Its effectiveness has been evaluated empirically in our experiments and found to yield robust performance across a range of settings.

\section{Experiments on Deletion Channels}\label{sec:experiments_synthetic}
\par This section describes the numerical evaluation of the \gls{npd} for deletion channels. For deletion channels, there is a trellis-based \gls{sc} decoder \cite{talPolarCodesDeletion2021} that allows us to benchmark the decoding performance of the \gls{npd}. Furthermore, known bounds on the channel capacity enable us to assess the \gls{npd}'s ability to estimate and optimize the \gls{mi} over input distributions.

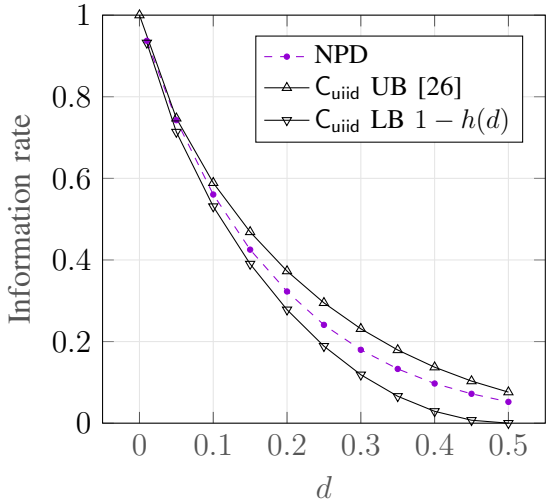
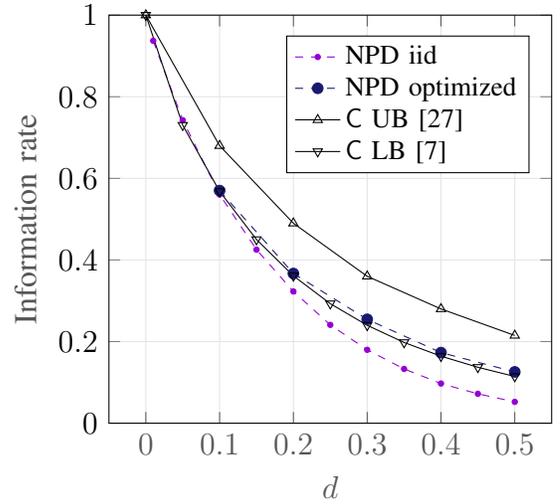
\begin{figure}[!b]
	\centering
	\begin{subfigure}{0.48\textwidth}
		\centering
		\begin{tikzpicture}

\begin{axis}[
width=2.94in,
height=2.76in,
x grid style={gainsboro229},
xlabel=\textcolor{dimgray85}{\(\displaystyle d\)},
xmajorgrids,
 xtick style={color=dimgray85},
 xtick={0, 0.1, 0.2, 0.3, 0.4, 0.5},
xticklabel style={color=dimgray85, /pgf/number format/.cd, fixed},
y grid style={gainsboro229},
ylabel=\textcolor{dimgray85}{Information rate},
ymajorgrids,
ymin=0.0, ymax=1.0,
ymajorticks=true,
legend cell align={left},
legend style={at={(.95,.95)}, anchor=north east, nodes={scale=0.85, transform shape}},
]
\addplot [dashed, darkviolet, mark=*, mark size=1, mark options={solid}]
table {%
0.01 	0.937170830280866
0.05   0.7425468436742877
0.1    	 0.5603768718504991
0.15	0.4252824538879956
0.2 	 0.32278065410582457
0.25	0.2409085501988796
0.3 	0.1797975736710078
0.35   0.13297737669792264
0.4		0.0971897698371128
0.45  0.0720237796220412
0.5    0.052244308623127456
};
\addlegendentry{NPD};
\addplot [solid, black, mark=triangle, mark size=2, mark options={solid}]
table {%
0.0   1.0  
0.05  0.747173  
0.10  0.589036  
0.15  0.468624  
0.20  0.372702  
0.25  0.294883  
0.30  0.231322  
0.35  0.179401  
0.40  0.137187  
0.45  0.10316   
0.50  0.076071  
};
\addlegendentry{$\sC_\mathsf{uiid}$ UB \cite{perniceMutualInformationUpper2024}};

\addplot [solid, black, mark=triangle, mark size=2, mark options={solid, rotate=180}]
table {%
0.01	   0.93212
0.05	0.7136030428840439
0.1	    0.5310044064107188
0.15	0.3901596952835995
0.2	    0.2780719051126377
0.25	0.18872187554086717
0.3	    0.1187091007693073
0.35	0.06593194462450891
0.4	    0.02904940554533142
0.45	0.007225546012191719
0.5	    0.0
};
\addlegendentry{$\sC_\mathsf{uiid}$  LB $1-h(d)$};

\end{axis}

\end{tikzpicture}
		\caption{\Gls{mi} estimated by the \gls{npd} for $N=128$, compared to known bounds: the \gls{iid} input upper bound $\mathsf{C}_{\text{iid}}$ \cite{perniceMutualInformationUpper2024} and the lower bound $1 - h(d)$.}
		\label{fig:del-mi-rate}
	\end{subfigure}
	\hfill
	\begin{subfigure}{0.48\textwidth}
		\centering
		\begin{tikzpicture}

\definecolor{chocolate2267451}{RGB}{226,74,51}
\definecolor{dimgray85}{RGB}{85,85,85}
\definecolor{gainsboro229}{RGB}{229,229,229}
\definecolor{darkblue}{RGB}{0, 0, 139}
\definecolor{darkred}{RGB}{139, 0, 0}
\definecolor{darkgreen}{RGB}{0, 100, 0}
\definecolor{darkgray}{RGB}{169, 169, 169}
\definecolor{midnightblue}{RGB}{25, 25, 112}
\definecolor{darkslategray}{RGB}{47, 79, 79}
\definecolor{darkolivegreen}{RGB}{85, 107, 47}
\definecolor{darkviolet}{RGB}{148, 0, 211}
\definecolor{darkorange}{RGB}{255, 140, 0}
\definecolor{darkmagenta}{RGB}{139, 0, 139}
\begin{axis}[
width=2.94in,
height=2.76in,
x grid style={gainsboro229},
xlabel=\textcolor{dimgray85}{\(\displaystyle d\)},
xmajorgrids,
xtick={0, 0.1, 0.2, 0.3, 0.4, 0.5},
 xtick style={color=dimgray85},
xticklabel style={color=dimgray85, /pgf/number format/.cd, fixed},
y grid style={gainsboro229},
ylabel=\textcolor{dimgray85}{Information rate},
ymajorgrids,
ymin=0.0, ymax=1.0,
ymajorticks=true,
legend cell align={left},
legend style={at={(.95,.95)}, anchor=north east, nodes={scale=0.85, transform shape}},
]
\addplot [dashed, darkviolet, mark=*, mark size=1, mark options={solid}]
table {%
	0.01 	0.937170830280866
	0.05   0.7425468436742877
	0.1    	 0.5603768718504991
	0.15	0.4252824538879956
	0.2 	 0.32278065410582457
	0.25	0.2409085501988796
	0.3 	0.1797975736710078
	0.35   0.13297737669792264
	0.4		0.0971897698371128
	0.45  0.0720237796220412
	0.5    0.052244308623127456
};
\addlegendentry{NPD iid};
\addplot [dashed, midnightblue, mark=*, mark size=2, mark options={solid}]
table {%

0.1    	 0.57
0.2 	0.3668
0.3 	0.2545
0.4     0.173
0.5		0.1257
};
\addlegendentry{NPD optimized};

\addplot [solid, black, mark=triangle, mark size=2, mark options={solid}]
table {%
0.0 1.0	   
0.1	    0.68
0.2	    0.49
0.3	    0.36
0.4	    0.28
0.5	    0.215
};
\addlegendentry{$\sC$  UB \cite{rahmatiUpperBoundsCapacity2015}};
\addplot [solid, black, mark=triangle, mark size=2, mark options={solid, rotate=180}]
table {%
	0.0 1.0
	0.05 0.7303 
	0.1 0.5681
	0.15 0.4501
	0.2 0.3611
	0.25 0.2929
	0.3 0.2399 
	0.35 0.1978
	0.4 0.1642
	0.45 0.1370 
	0.5 0.1145 
};
\addlegendentry{$\sC$ LB \cite{castiglioneTrellisBasedLower2015}};
\end{axis}

\end{tikzpicture}
		\caption{Optimized \gls{mi} estimated by the \gls{npd}, compared with known lower and upper bounds on the deletion channel capacity.}
		\label{fig:del-opt-mi-rate}
	\end{subfigure}
	\caption{Comparison of \gls{npd}-based \gls{mi} estimates with theoretical bounds for the deletion channel.}
	\label{fig:del-mi-all}
\end{figure}


\subsubsection{Setup}

\par In all experiments, the \gls{npd} was trained on $5\cdot 10^6$ independent input-output pairs $(\mathbf{X},\mathbf{Y})$. For each block length $N \in \{32, 64, 128, 256, 512\}$, a separate \gls{npd} model was trained. After training, the parameters of each \gls{npd} were fixed and used for decoding.

\par Additionally, we evaluated the \gls{npd} under \gls{scl} decoding, following \cite{talPolarCodesDeletion2021,aharoniDatadrivenNeuralPolar2024}. However, due to the high computational complexity of the trellis-based decoder, we were unable to perform a comparable evaluation with \gls{scl} decoding for the trellis decoder.

\subsubsection{Information Rates}

\par This section demonstrates the utility of using the \gls{npd} to estimate and optimize the \gls{mi} between channel inputs and outputs. For this purpose, we follow the methodology for \gls{mi} estimation described in Section~\ref{sec:npd_mi_est} and the methodology for \gls{mi} optimization described in Section~\ref{sec:npd_mi_opt}.

\par In Figures~\ref{fig:del-mi-rate} and~\ref{fig:del-opt-mi-rate}, the \gls{npd} estimates of the \gls{mi} are shown for varying deletion probabilities $d \in \{0.1, 0.2, 0.3, 0.4, 0.5\}$, closely matching known theoretical bounds. In Figure~\ref{fig:del-mi-rate}, the \gls{npd} estimate for \gls{iid} inputs with $N=128$ lies between the known upper and lower bounds. Figure~\ref{fig:del-opt-mi-rate} demonstrates that optimizing the input distribution further improves the \gls{mi} estimates, particularly at high deletion rates. These results highlight the effectiveness of the \gls{npd}, especially when input optimization is employed.

\begin{remark}
The \gls{npd}-based \gls{mi} estimates presented here are not rigorous bounds on the channel capacity. They are computed as the difference of two numerical upper bounds on entropy, and they are obtained at finite block lengths, not in the asymptotic regime. The known lower and upper bounds shown are used solely as benchmarks to evaluate the proposed method, rather than to claim precise capacity estimation.
\end{remark}
\begin{figure}[!h]
	\centering
	\begin{subfigure}{0.45\textwidth}
		\centering
		\centering
		    \begin{tikzpicture}
    	\definecolor{chocolate2267451}{RGB}{226,74,51}
    	\definecolor{dimgray85}{RGB}{85,85,85}
    	\definecolor{gainsboro229}{RGB}{229,229,229}
    	\definecolor{darkblue}{RGB}{0, 0, 139}
    	\definecolor{darkred}{RGB}{139, 0, 0}
    	\definecolor{darkgreen}{RGB}{0, 100, 0}
    	\definecolor{darkgray}{RGB}{169, 169, 169}
    	\definecolor{midnightblue}{RGB}{25, 25, 112}
    	\definecolor{darkslategray}{RGB}{47, 79, 79}
    	\definecolor{darkolivegreen}{RGB}{85, 107, 47}
    	\definecolor{darkviolet}{RGB}{148, 0, 211}
    	\definecolor{darkorange}{RGB}{255, 140, 0}
    	\definecolor{darkmagenta}{RGB}{139, 0, 139}
	\begin{axis}[
width=2.94in,
height=2.76in,
		xlabel=\textcolor{dimgray85}{$\log_2 N$},
		ylabel=\textcolor{dimgray85}{Decoder speed (blocks/sec)},
ylabel style={
	at={(axis description cs:-0.01,0.5)}, 
	anchor=south
},
		symbolic x coords={5, 6, 7, 8, 9},
		xtick=data,
		yticklabels={},
		ybar=2pt,
		bar width=10pt,
		ymax=3750,
		legend pos=north east,
		legend style={font=\small},
		enlarge x limits=0.12,
		y tick label style={rotate=90},
		every node near coord/.append style={font=\small, rotate=90, anchor=west}, 
		nodes near coords={\pgfmathprintnumber[fixed, precision=3]{\pgfplotspointmeta}}, 
		nodes near coords align={vertical},
		]
		\addplot+ [bar shift=0pt] coordinates {(5, 30.16) (6, 10.74) (7, 2.67) (8, 0.92) (9, 0.16)};
		\addplot+ [bar shift=-12pt] coordinates {(5, 8.39) (6, 1.83) (7, 0.26) (8, 0.02) (9, 0.001)};
		\addplot+ [bar shift=12pt] coordinates {(5, 2907) (6, 1470) (7, 757) (8, 367) (9, 185)};
		\legend{SCT $\delta=0.01$, SCT $\delta=0.1$, NPD}
	\end{axis}
\end{tikzpicture}
		\caption{}
		\label{fig:complexity}
	\end{subfigure}
	\hfill
	\begin{subfigure}{0.5\textwidth}
		\centering
\begin{tikzpicture}

\definecolor{chocolate2267451}{RGB}{226,74,51}
\definecolor{dimgray85}{RGB}{85,85,85}
\definecolor{gainsboro229}{RGB}{229,229,229}
\definecolor{darkblue}{RGB}{0, 0, 139}
\definecolor{darkred}{RGB}{139, 0, 0}
\definecolor{darkgreen}{RGB}{0, 100, 0}
\definecolor{darkgray}{RGB}{169, 169, 169}
\definecolor{midnightblue}{RGB}{25, 25, 112}
\definecolor{darkslategray}{RGB}{47, 79, 79}
\definecolor{darkolivegreen}{RGB}{85, 107, 47}
\definecolor{darkviolet}{RGB}{148, 0, 211}
\definecolor{darkorange}{RGB}{255, 140, 0}
\definecolor{darkmagenta}{RGB}{139, 0, 139}
	\begin{axis}[
width=2.94in,
height=2.76in,
		xlabel=\textcolor{dimgray85}{$\log_2 N$},
		symbolic x coords={5, 6, 7, 8, 9},
		xtick=data,
		ybar=2pt,
		bar width=10pt,
		ymax=1.0,
		legend pos=north east,
		legend style={font=\small},
		axis y line*=right,
		yticklabels={},
		every node near coord/.append style={font=\small, rotate=90, anchor=west}, 
		nodes near coords={\pgfmathprintnumber[fixed, precision=3]{\pgfplotspointmeta}}, 
		nodes near coords align={vertical},
		]
		\addplot+[opacity=0.5] coordinates {(5, 0.687) (6, 0.703) (7, 0.757) (8, 0.768) (9, 0.775)};
		\addplot+[opacity=0.5] coordinates {(5, 0.25) (6, 0.31) (7, 0.328) (8, 0.34) (9, 0.351)};
	\end{axis}
	\begin{axis}[
width=2.94in,
height=2.76in,
		ylabel=\textcolor{dimgray85}{FER},
		xlabel={}, 
		xtick=\empty,
		ytick pos=left,
		ymin=0,
		ymax=0.1,
		yticklabel style={/pgf/number format/fixed,/pgf/number format/precision=2}, 
		legend pos=north east,
		legend style={font=\small}
		]
		\addplot[only marks, mark=*, color=blue, mark options={fill=blue, draw=blue, line width=0.5pt, scale=1}] coordinates {(5, 0.084) (6, 0.088) (7, 0.084) (8, 0.084) (9, 0.079)};
		\addplot[only marks, mark=*, mark options={fill=white, draw=blue, line width=0.5pt, scale=1}] coordinates {(5, 0.06) (6, 0.058) (7, 0.056) (8, 0.057) (9, 0.055)};
		\addplot[only marks, mark=*, color=red, line width=1pt, fill=none] coordinates {(5, 0.085) (6, 0.082) (7, 0.079) (8, 0.082) };
		\addplot[only marks, mark=*, mark options={fill=white, draw=red, line width=0.5pt, scale=1}] coordinates {(5, 0.065) (6, 0.062) (7, 0.059) (8, 0.058) (9, 0.057)};
	\end{axis}
\end{tikzpicture}
		\caption{}
		\label{fig:fer-sc}
	\end{subfigure}
	\caption{(a) Comparison of decoder speed of the \gls{npd} and the trellis-based for deletion channels. (b) Comparison of the SCT and the NPD for various block lengths. The bar plot shows the code rate per block length. The marks shows the attained \glspl{fer}. Blue/red plots correspond to $d=0.01,0.1$, respectively. The input distribution is \gls{iid}.}
\end{figure}
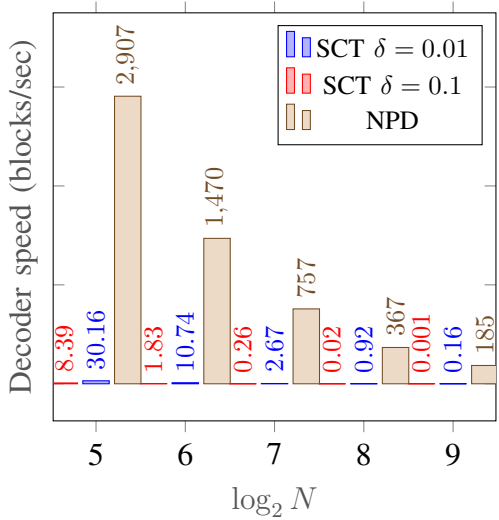
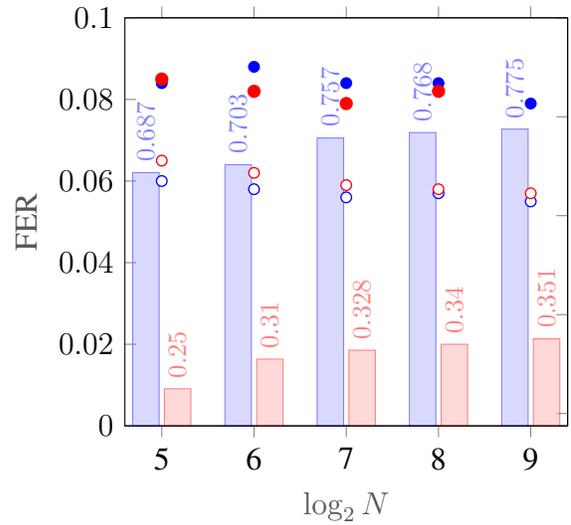

\begin{figure}[!h]
	\centering
	\begin{subfigure}[b]{0.48\textwidth}
		\centering
		    \begin{tikzpicture}
\definecolor{chocolate2267451}{RGB}{226,74,51}
\definecolor{dimgray85}{RGB}{85,85,85}
\definecolor{gainsboro229}{RGB}{229,229,229}
\definecolor{darkblue}{RGB}{0, 0, 139}
\definecolor{darkred}{RGB}{139, 0, 0}
\definecolor{darkgreen}{RGB}{0, 100, 0}
\definecolor{darkgray}{RGB}{169, 169, 169}
\definecolor{midnightblue}{RGB}{25, 25, 112}
\definecolor{darkslategray}{RGB}{47, 79, 79}
\definecolor{darkolivegreen}{RGB}{85, 107, 47}
\definecolor{darkviolet}{RGB}{148, 0, 211}
\definecolor{darkorange}{RGB}{255, 140, 0}
\definecolor{darkmagenta}{RGB}{139, 0, 139}
	\begin{axis}[
width=2.94in,
height=2.76in,
		ylabel=\textcolor{dimgray85}{Rate},
		symbolic x coords={5, 6, 7, 8, 9},
		xtick=data,
		ybar=2pt,
		bar width=10pt,
		ymax=0.8,
		axis y line*=right,
		xlabel={}, 
		xtick=\empty,
		ytick=\empty,
		ytick pos=right,
		y tick label style={rotate=90},
		every node near coord/.append style={font=\small, rotate=90, anchor=west}, 
		nodes near coords={\pgfmathprintnumber[fixed, precision=3]{\pgfplotspointmeta}}, 
		nodes near coords align={vertical},
		]
		\addplot+[opacity=0.35] coordinates {(5, 0.687) (6, 0.703) (7, 0.757) (8, 0.768) (9, 0.775)};
	\end{axis}
	\begin{axis}[
width=2.94in,
height=2.76in,
		xlabel=\textcolor{dimgray85}{$\log_2 N$},
		ylabel=\textcolor{dimgray85}{FER},
		symbolic x coords={5, 6, 7, 8, 9},
		xtick=data,
		ymax=0.1,
		legend style={at={(0.5,1.0)},
			font=\small,
			anchor=south,},
		legend columns=2,
		ytick pos=left,
		yticklabel style={/pgf/number format/fixed,/pgf/number format/precision=2}, 
		nodes near coords align={vertical},
		]
		\addplot[mark=*, color=black, mark options={fill=black, draw=black, line width=0.5pt, scale=1}] coordinates {(5, 0.084) (6, 0.088) (7, 0.084) (8, 0.084) (9, 0.079)};
		\addplot[mark=*, color=darkblue, mark options={fill=white, draw=darkblue, line width=0.1pt, scale=1}] coordinates {(5, 0.06) (6, 0.058) (7, 0.056) (8, 0.057) (9, 0.055)};
		\addplot[mark=square*, color=darkred, mark options={fill=white, draw=darkred, line width=0.1pt, scale=1}] coordinates {(5, 0.026) (6, 0.025) (7, 0.027) (8, 0.029) (9, 0.0296)};
		\addplot[mark=triangle*, color=darkgreen, mark options={fill=white, draw=darkgreen, line width=0.1pt, scale=1}] coordinates {(5, 0.023) (6, 0.024) (7, 0.0206) (8, 0.024) (9, 0.0233)};
		\addplot[mark=triangle*, color=darkorange, mark options={rotate=180,fill=white, draw=darkorange, line width=0.1pt, scale=1}] coordinates {(5, 0.022) (6, 0.0237) (7, 0.021) (8, 0.023) (9, 0.0229)};
		\legend{SCT $L=1$, NPD $L=1$, NPD $L=2$,NPD $L=4$,NPD $L=8$,}
	\end{axis}
\end{tikzpicture}
		\caption{}
		\label{fig:fer-scl-01}
	\end{subfigure}
	\hfill
	\begin{subfigure}[b]{0.48\textwidth}
		\centering
		    \begin{tikzpicture}
\definecolor{chocolate2267451}{RGB}{226,74,51}
\definecolor{dimgray85}{RGB}{85,85,85}
\definecolor{gainsboro229}{RGB}{229,229,229}
\definecolor{darkblue}{RGB}{0, 0, 139}
\definecolor{darkred}{RGB}{139, 0, 0}
\definecolor{darkgreen}{RGB}{0, 100, 0}
\definecolor{darkgray}{RGB}{169, 169, 169}
\definecolor{midnightblue}{RGB}{25, 25, 112}
\definecolor{darkslategray}{RGB}{47, 79, 79}
\definecolor{darkolivegreen}{RGB}{85, 107, 47}
\definecolor{darkviolet}{RGB}{148, 0, 211}
\definecolor{darkorange}{RGB}{255, 140, 0}
\definecolor{darkmagenta}{RGB}{139, 0, 139}
\begin{axis}[
width=2.94in,
height=2.76in,
		ylabel=\textcolor{dimgray85}{Rate},
		symbolic x coords={5, 6, 7, 8, 9},
		xtick=data,
		ybar=2pt,
		bar width=10pt,
		ymax=0.39,
		axis y line*=right,
		xlabel={}, 
		xtick=\empty,
		ytick=\empty,
		ytick pos=right,
		y tick label style={rotate=90},
		every node near coord/.append style={font=\small, rotate=90, anchor=west}, 
		nodes near coords={\pgfmathprintnumber[fixed, precision=3]{\pgfplotspointmeta}}, 
		nodes near coords align={vertical},
		]
		\addplot+[opacity=0.0] coordinates {(5, 0.687) (6, 0.703) (7, 0.757) (8, 0.768) (9, 0.775)};
		\addplot+[bar shift=-1pt, opacity=0.35] coordinates {(5, 0.25) (6, 0.31) (7, 0.328) (8, 0.34) (9, 0.351)};
	\end{axis}
	\begin{axis}[
width=2.94in,
height=2.76in,
		xlabel=\textcolor{dimgray85}{$\log_2 N$},
		ylabel=\textcolor{dimgray85}{ FER },
		symbolic x coords={5, 6, 7, 8, 9},
		xtick=data,
		ymax=0.1,
		legend style={at={(0.5,1.0)},
			font=\small,
			anchor=south,},
		legend columns=2,
		ytick pos=left,
		yticklabel style={/pgf/number format/fixed,/pgf/number format/precision=2}, 
		nodes near coords align={vertical},
		]
		\addplot[mark=*, color=black, mark options={fill=black, draw=black, line width=0.5pt, scale=1}] coordinates {(5, 0.084) (6, 0.088) (7, 0.084) (8, 0.084) (9, 0.079)};
		\addplot[mark=*, color=darkblue, mark options={fill=white, draw=darkblue, line width=0.1pt, scale=1}] coordinates {(5, 0.06) (6, 0.058) (7, 0.056) (8, 0.057) (9, 0.055)};
		\addplot[mark=square*, color=darkred, mark options={fill=white, draw=darkred, line width=0.1pt, scale=1}] coordinates {(5, 0.026) (6, 0.025) (7, 0.027) (8, 0.029) (9, 0.0296)};
		\addplot[mark=triangle*, color=darkgreen, mark options={fill=white, draw=darkgreen, line width=0.1pt, scale=1}] coordinates {(5, 0.023) (6, 0.024) (7, 0.0206) (8, 0.024) (9, 0.0233)};
		\addplot[mark=triangle*, color=darkorange, mark options={rotate=180,fill=white, draw=darkorange, line width=0.1pt, scale=1}] coordinates {(5, 0.022) (6, 0.0237) (7, 0.021) (8, 0.023) (9, 0.0229)};
		\legend{SCT $L=1$, NPD $L=1$, NPD $L=2$,NPD $L=4$,NPD $L=8$,}
	\end{axis}
\end{tikzpicture}
		\caption{}
		\label{fig:fer-scl-1}
	\end{subfigure}
	\caption{\Glspl{fer} attained by SCL decoding of the \gls{npd} for  $d=0.01,0.1$ in (a),(b), respectively. The bar plot shows the information rate per block length. The input distribution is \gls{iid}.}
	
\end{figure}
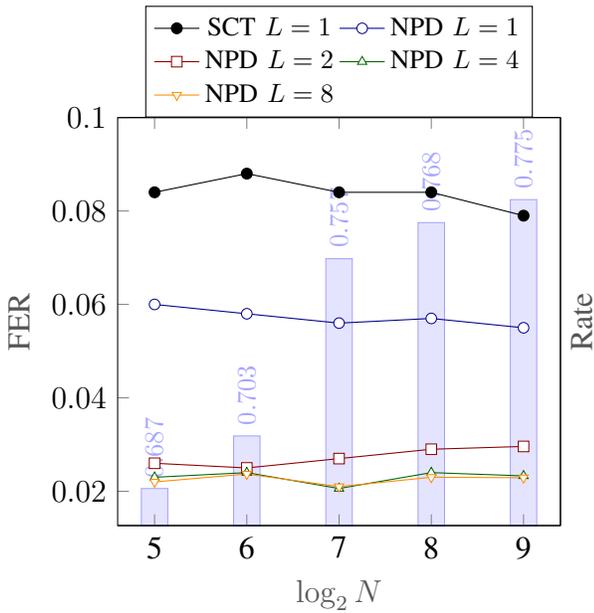
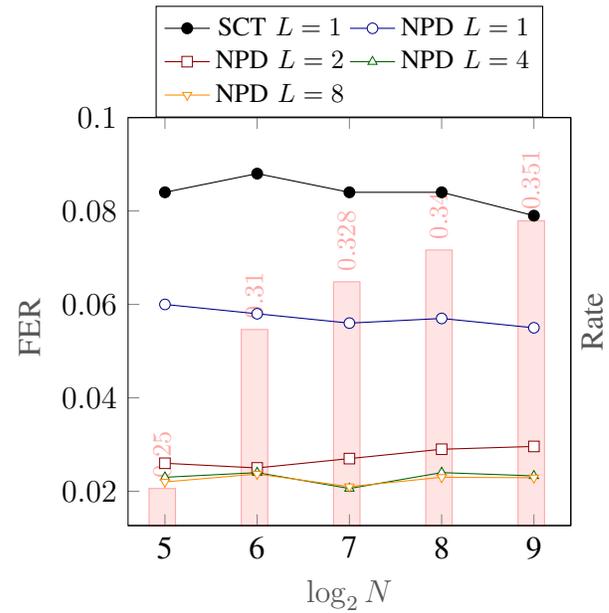

\subsubsection{Computational Complexity}

\par Figure~\ref{fig:complexity} compares the decoding speed, measured in blocks per second, of the \gls{npd} and the trellis-based decoder from \cite{talPolarCodesDeletion2021}. Results are shown for both $d = 0.01$ and $d = 0.1$, since the computational complexity of trellis-based \gls{sc} decoding depends on the number of deletions. In contrast, the computational complexity of the \gls{npd} depends only on the parametrization of the neural networks and is independent of the channel realization. Thus, the \gls{npd}'s decoding speed remains the same for both values of $d$.

\par We note that the comparison is not entirely fair, as the \gls{npd} was executed on a GPU while the trellis-based decoder was executed on a CPU. Nevertheless, the decoding speed gap is sufficiently large that the advantage of the \gls{npd} remains clear. The results convincingly demonstrate that the \gls{npd} achieves significantly faster decoding across all tested configurations. This computational advantage is particularly critical for enabling more complex decoding strategies. In the next section, we demonstrate that SCL decoding could be applied to the \gls{npd} implementation, which would have been computationally infeasible with the trellis-based decoder.


\subsubsection{Frame Error Rate of SC Decoding}

\par Figure~\ref{fig:fer-sc} compares the trellis-based decoder and the \gls{npd} by illustrating the \glspl{fer} attained by both decoders across various block lengths. The blue and red bars correspond to $d=0.01$ and $d=0.1$, respectively. The codes were designed to achieve a target \gls{fer} around $0.1$, and the bars indicate the information rate used per block length. Solid markers represent the \glspl{fer} attained by the trellis-based decoder, while hollow markers represent the \glspl{fer} attained by the \gls{npd}. 

\par The results show that the \gls{npd} achieves slightly better \glspl{fer} than the trellis-based decoder. This improvement is somewhat surprising, as theoretically, both decoders should perform similarly. We conjecture that the difference is due to numerical errors in the implementation of the trellis-based decoder in \cite{talPolarCodesDeletion2021}.

\subsubsection{Frame Error Rate of SCL Decoding}

\par Figures~\ref{fig:fer-scl-01} and~\ref{fig:fer-scl-1} show the performance of \gls{scl} decoding for $d=0.01$ and $d=0.1$, respectively. Only the \gls{npd} results are shown, since the trellis-based decoder already exhibits prohibitively high running time under \gls{sc} decoding alone, making \gls{scl} decoding infeasible. 

\par As expected, incorporating \gls{scl} decoding with the \gls{npd} leads to a significant reduction in the attained \glspl{fer}. These results further demonstrate the computational advantage of the \gls{npd}, enabling advanced decoding strategies that would otherwise be impractical for deletion channels.

\subsection{Numerical Results for \gls{ids} Channels}

\par This section presents empirical results on \gls{ids} channels. We consider uniform and \gls{iid} input distributions and evaluate the performance on instances of \gls{ids} channels with various insertion, deletion, and substitution probabilities. Since in the \gls{ids} channel the output sequence can be either longer or shorter than the input sequence, the attention-based \gls{nn} is deployed.

\par To validate the performance of the \gls{npd}, we trained a separate \gls{npd} for each combination of $(i,d,s)$ values for which bounds are available in the literature. We refer to \cite{fertonaniBoundsCapacityChannels2011} for known lower and upper bounds on the capacity of \gls{ids} channels. Table~\ref{tab:ids_results} compares the estimated \gls{mi} values obtained by the \gls{npd} with the theoretical bounds.

\par As shown in Table~\ref{tab:ids_results}, the \gls{npd}'s estimates consistently lie within the known bounds for all tested parameters. This result is noteworthy, as the \gls{npd} estimates were generated using a uniform and \gls{iid} input distribution, whereas the bounds pertain to the channel capacity, which may be achieved by a nonuniform input distribution. We speculate that this strong performance is due to two factors: (i) at low error probabilities, the capacity-achieving input distribution is close to uniform, and (ii) at higher error probabilities, the known bounds become looser.

\begin{table}[!h]
\centering
\begin{tabular}{|ccc|c|cc|}
\hline
$d$ & $i$ & $s$ & NPD & LB & UB \\
\hline
0.01 & 0.00 & 0.01 & \textbf{0.856} & 0.842 & 0.886 \\
0.01 & 0.00 & 0.03 & \textbf{0.745} & 0.732 & 0.776 \\
0.01 & 0.00 & 0.10 & \textbf{0.478} & 0.466 & 0.510 \\
0.05 & 0.00 & 0.01 & \textbf{0.662} & 0.653 & 0.767 \\
0.05 & 0.00 & 0.03 & \textbf{0.560} & 0.555 & 0.669 \\
0.05 & 0.00 & 0.10 & \textbf{0.332} & 0.321 & 0.435 \\
0.10 & 0.00 & 0.01 & \textbf{0.495} & 0.492 & 0.644 \\
0.10 & 0.00 & 0.03 & \textbf{0.410} & 0.408 & 0.560 \\
0.10 & 0.00 & 0.10 & \textbf{0.220} & 0.211 & 0.363 \\
0.01 & 0.01 & 0.01 & \textbf{0.781} & 0.766 & 0.863 \\
0.01 & 0.03 & 0.01 & \textbf{0.678} & 0.661 & 0.808 \\
0.01 & 0.10 & 0.01 & \textbf{0.432} & 0.412 & 0.642 \\
0.03 & 0.01 & 0.01 & \textbf{0.681} & 0.662 & 0.808 \\
0.03 & 0.03 & 0.01 & \textbf{0.578} & 0.564 & 0.750 \\
0.03 & 0.10 & 0.01 & \textbf{0.369} & 0.329 & 0.583 \\
0.10 & 0.01 & 0.01 & \textbf{0.434} & 0.419 & 0.649 \\
0.10 & 0.03 & 0.01 & \textbf{0.366} & 0.335 & 0.589 \\
0.10 & 0.10 & 0.01 & \textbf{0.216} & 0.139 & 0.438 \\
\hline
\end{tabular}
\caption{Comparison of \gls{npd}-estimated information rates with lower and upper bounds from \cite{fertonaniBoundsCapacityChannels2011}.}
\label{tab:ids_results}
\end{table}

\section{Experiments on DNA Data Storage Systems}\label{sec:experiments_dna}

\par This section presents experimental results on realistic DNA data storage systems. It begins with experiments on synthetic data, where the DNA storage channel is modeled as an \gls{ids} channel over a $4$-ary alphabet with multiple independent traces. It then proceeds to experiments on real-world DNA data published in \cite{srinivasavaradhanTrellisBMACoded2024}.

\par In both settings, the \gls{npd} is used to implement the inner code, operating on DNA strands of $128$ $4$-ary symbols. Furthermore, we demonstrate how the \gls{npd} can be scaled to implement an outer code by concatenating multiple DNA strands. This enables the simulation of sequences of lengths up to $524{,}288$ bits on a single GPU.

\subsection{Synthetic DNA Experiments}

\par This subsection validates the performance of the \gls{npd} on synthetic DNA storage data before proceeding to experiments on real DNA sequencing data.

\subsubsection{DNA Data Storage Channel Model}

\par Let $\mathbf{x} \in \mathbb{F}_2^N$ denote the binary input and $\mathbf{y} \in \mathbb{F}_4^\ast$ denote the channel output, where $\mathbb{F}_4^\ast = \bigcup_{n\in\mathbb{N}} \mathbb{F}_4^n$. The DNA storage channel model follows these steps: given $\mathbf{x}$, it is first mapped to a quaternary alphabet. Then, the number of observed outputs is sampled as $K \sim \operatorname{Poisson}(\lambda)$, where $\lambda > 0$, and $K$ independent outputs $\mathbf{y}^{1:K} = (\mathbf{y}^1, \dots, \mathbf{y}^K)$ are drawn, each through an independent instance of an \gls{ids} channel.

\subsubsection{Training the NPD for the DNA Storage Channel Model}

\par For the DNA storage channel model, the \gls{npd} uses the attention-based embedding described in Definition~\ref{def:att_emb}. In this setting, we set $L_\text{max} = \lceil 1.1  \frac{N}{2} \rceil$ and the output alphabet $\Sigma = \{0,1,2,3\}$. To accommodate multiple traces, the \gls{npd} is trained using the embedding aggregation defined in \eqref{eqn:multiple_emb}. Apart from these changes, the training procedure follows the same method described in Section~\ref{sec:npd_mi_est}.
The \gls{npd} is trained for $N = 256$, corresponding to DNA strands of length $128$ symbols. After training, the \gls{npd} parameters are fixed and used for all subsequent evaluations.

\begin{figure}[t!]
	\centering
	\begin{tikzpicture}

	\begin{axis}[
width=2.94in,
height=2.76in,
		x grid style={gainsboro229},
		xlabel=\textcolor{dimgray85}{Number of Traces },
		xmajorgrids,
		xticklabel style={color=dimgray85},
		y grid style={gainsboro229},
		ylabel=\textcolor{dimgray85}{FER},
		ymajorgrids,
		ymode=log,  
		log basis y={10},
		xtick={1,2,3,4,5,6,7,8,9,10, 12}, 
		xticklabels={1,2,3,4,5,6,7,8,9,10,$K$}, 
		ytick={2e-2, 5e-2, 7e-2, 1e-1,1.5e-1, 2e-1, 2.5e-1}, 
		log ticks with fixed point, 
		ymax=0.1,
		scaled y ticks=manual:{}
		yminorgrids=false,  
		yminorticks=false,
		ytick style={draw=none},
		legend cell align={left},
		legend style={at={(-0.175,0.1)}, anchor=east, nodes={scale=0.85, transform shape}},
		]
		\addplot [only marks, darkviolet, mark=*, mark size=2, mark options={solid}]
		table {%
			1	 0.0628
			2 	0.0632
			3   0.0538
			4  0.0594
			5  0.0598
			6  0.058
			7   0.0576
			8  0.0426
			9  0.0406
			10 0.0222
			12 0.059100
		};
		\addlegendentry{SC};
		\addplot [only marks, midnightblue, mark=*, mark size=2, mark options={solid}]
		table {%
			1   0.03
			2   0.0278
			3   0.0234
			4  0.0326
			5  0.0462
			6  0.0488
			7   0.0552
			8   0.0444
			9   0.0386
			10 0.0238
			12 0.039000
		};
		\addlegendentry{SCL 8};
	\end{axis}
	\begin{axis}[
width=2.94in,
height=2.76in,
		axis y line*=right,              
		axis x line=none,                
		ybar=0pt,                        
		bar width=6pt,                  
		ylabel=\textcolor{dimgray85}{Information Rate},
		ymin=0.3,
		ymax=1.05,
		ytick={0, 0.2, 0.4, 0.6, 0.8, 1.0},
		xtick={1,2,3,4,5,6,7,8,9,10, 12}, 
		xticklabels={},                 
		legend style={at={(1.15,0.1)}, anchor=west, nodes={scale=0.85, transform shape}},
		]
		
		\addplot+[fill=darkblue, fill opacity=0.25, draw=black, line width=0.5pt] coordinates {
			(1, 0.71537)
			(2, 0.87341)
			(3,0.94711)
			(4, 0.97637)
			(5, 0.98888)
			(6, 0.99451)
			(7, 0.99716)
			(8, 0.99843)
			(9, 0.99911)
			(10,0.99943)
			(12,0.942693)
		};
			\addlegendentry{MI};
		\addplot+[fill=darkmagenta, fill opacity=0.25, draw=black, line width=0.5pt] coordinates {
			(1, 0.335)
			(2, 0.605)
			(3,0.746)
			(4, 0.878)
			(5, 0.921)
			(6, 0.9686)
			(7, 0.9849)
			(8, 0.9963)
			(9, 1.)
			(10,1.)
			(12,0.64)
		};
					\addlegendentry{Code rate};
	\end{axis}
	
\end{tikzpicture}
	\caption{Performance of the \gls{npd} when decoding blocks of $128$ DNA symbols. Results are shown as a function of the number of observed traces. Each point represents $10{,}000$ Monte Carlo samples. The column labeled $K$ shows results for $K \sim \operatorname{Poisson}(5)$.}
	\label{fig:dna-traces}
\end{figure}
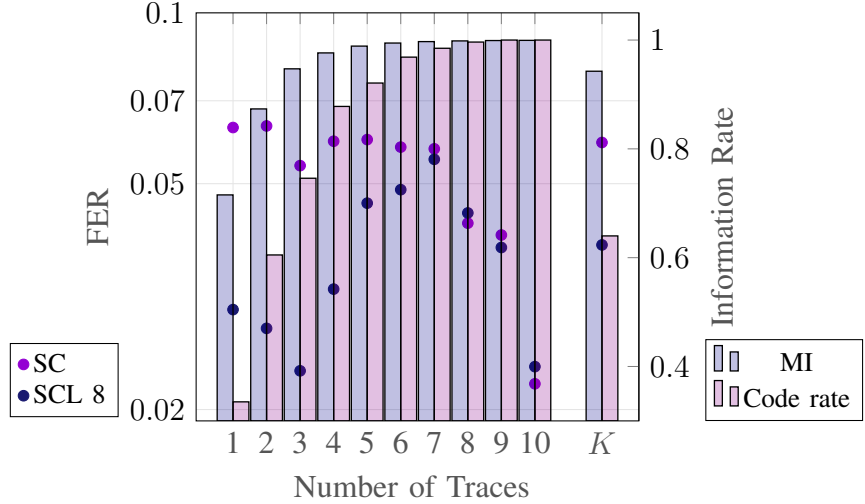
\subsubsection{SC Decoding Performance of the NPD on a Single Input Strand}
\par We first evaluate the performance of the \gls{npd} for each number of observed traces $K = 1, \dotsc, 10$. We do not test $K=0$ since, in that case, the input distribution is uniform and \gls{iid}, and the synthetic channel is pure noise. Figure~\ref{fig:dna-traces} shows the decoding performance of the \gls{npd} for each value of $K$ as well as for the case where $K \sim \operatorname{Poisson}(5)$. The figure presents the achieved \glspl{fer}, the estimated \gls{mi}, and the actual code rate required to attain \glspl{fer} around $0.07$. We also show the decoding performance under \gls{scl} decoding. In the figure, bar plots illustrate the information and code rates, while scatter plots show the \glspl{fer} for both \gls{sc} and \gls{scl} decoding.

\subsubsection{SC Decoding of Information Encoded on Multiple Input Strands}
\par The results in the previous section reveal substantial variability between the average information rate and the actual code rate of the \gls{npd} when $K$ is small. Furthermore, when input strands are completely lost, large deviations occur between the average and maximum decoding error. To mitigate these issues, we propose a scheme to decode information encoded across multiple input strands, motivated by methods presented in \cite{talPolarCodesDeletion2021}.

\par Let $N_0$ denote the number of DNA strands used to store the data. The polar coding scheme encodes $\mathbf{u} \in \mathbb{F}_2^{NN_0}$ into $\mathbf{x} = \mathbf{u} G_{NN_0}$, where $G_{NN_0}$ is the polar transform matrix. The resulting codeword $\mathbf{x}$ is then divided into $N_0$ length-$N$ sequences $\{\mathbf{x}^{(i)}\}_{i=1}^{N_0}$, each stored independently using the DNA storage channel model.

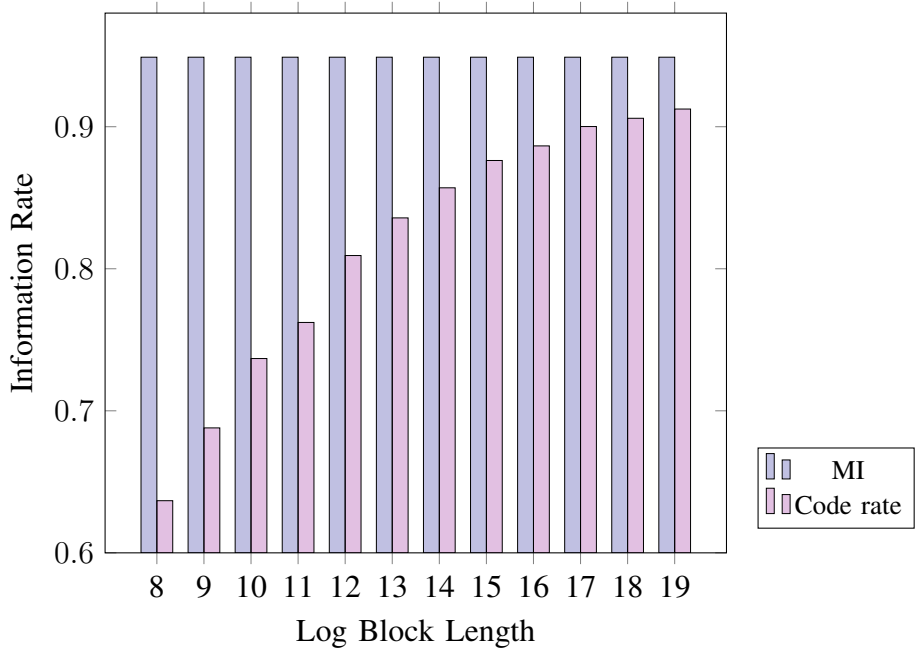
\begin{figure}[h!]
	\centering
\begin{tikzpicture}

\definecolor{chocolate2267451}{RGB}{226,74,51}
\definecolor{dimgray85}{RGB}{85,85,85}
\definecolor{gainsboro229}{RGB}{229,229,229}
\definecolor{darkblue}{RGB}{0, 0, 139}
\definecolor{darkred}{RGB}{139, 0, 0}
\definecolor{darkgreen}{RGB}{0, 100, 0}
\definecolor{darkgray}{RGB}{169, 169, 169}
\definecolor{midnightblue}{RGB}{25, 25, 112}
\definecolor{darkslategray}{RGB}{47, 79, 79}
\definecolor{darkolivegreen}{RGB}{85, 107, 47}
\definecolor{darkviolet}{RGB}{148, 0, 211}
\definecolor{darkorange}{RGB}{255, 140, 0}
\definecolor{darkmagenta}{RGB}{139, 0, 139}
\begin{axis}[
width=3.875in,
	height=3.45in,
	xmode=log,
	log basis x={2},
	ybar=0pt,                        
	bar width=6pt,                  
	ylabel={Information Rate},
	xlabel={Log Block Length},
	ymin=0.6,
	ymax=0.98,
	xtick={256,512,1024,2048,4096,8192,16384,32768,65536,131072,262144,524288}, 
	xticklabels={8,9,10,11,12,13,14,15,16,17,18,19},
	legend style={at={(1.05,0.12)}, anchor=west, nodes={scale=0.85, transform shape}},
	]
	
	\addplot+[fill=darkblue, fill opacity=0.25, draw=black, line width=0.25pt] coordinates {
		(256,0.948945)
		(512,0.948945)
		(1024,0.948945)
		(2048,0.948945)
		(4096,0.948945)
		(8192, 0.948945)		
		(16384, 0.948945)
		(32768, 0.948945)
		(65536, 0.948945)
		(131072, 0.948945)
		(262144, 0.948945)
		(524288, 0.948945)
	};
\addlegendentry{MI};
	\addplot+[fill=darkmagenta, fill opacity=0.25, draw=black, line width=0.25pt] coordinates {
	(256,0.636719)
	(512,0.687969)
	(1024,0.736797)
	(2048,0.762207)
	(4096,0.809316)
	(8192, 0.835811)
	(16384, 0.856995)
	(32768, 0.876230)
	(65539, 0.8865)
	(131072, 0.900126)
	(262144, 0.905945)
	(524288, 0.91245)
};
\addlegendentry{Code rate};
\end{axis}

\end{tikzpicture}
	\caption{Performance of the \gls{npd} when decoding information stored across multiple DNA strands. The number of traces follows a $\operatorname{Poisson}(5)$ distribution. Results are shown as a function of the number of concatenated strands $N_0$. Each point is based on $1000$ Monte Carlo samples.}
	\label{fig:dna-multistrand}
\end{figure}

\par The \gls{sc} decoder for this model is a hybrid between the \gls{npd} decoder and the standard \gls{sc} decoder for memoryless channels. At recursion depths corresponding to the first $\log_2 N$ stages, the decoder uses the functions $E_\theta$, $F_\theta$, $G_\theta$, and $H_\theta$ of the \gls{npd}. At recursion depths corresponding to the final $\log_2 N_0$ stages, where independent strands are combined, the decoder first projects the embeddings $e \in \mathbb{R}^d$ into log-likelihood ratios using $H_\theta$. It then switches to using the standard memoryless-channel functions $F$, $G$, and $H$ to complete the decoding.

\par Figure~\ref{fig:dna-multistrand} shows the performance of this multi-strand decoding scheme for various values of $N_0$. The results indicate that polar codes of length $NN_0$ can achieve \glspl{ber} around $0.02$. The figure reports both the estimated information rate and the actual code rate of the polar code when decoded under \gls{sc} decoding.

\subsection{Real DNA Nanopore Dataset}
\par This section demonstrates the performance of the proposed method on ``real-world'' DNA sequencing data. We apply the \gls{npd} framework to DNA data sequenced using Nanopore technology, which is known to operate in a higher noise regime compared to other sequencing methods.

\par The dataset contains $10{,}000$ DNA read-write examples. Each example includes a single input sequence consisting of $110$ $4$-ary symbols (representing the written sequence) and a cluster of reads, where each read has variable length. To evaluate our method, we partition the data into two groups: (i) a validation set containing 25\% of the data ($2500$ samples) used for training and tuning the \gls{npd}, and (ii) a test set containing the remaining 75\% of the data, used solely for performance evaluation.

\subsubsection{Training on Real DNA Data from Limited Samples}
\par The training procedure consists of three stages. First, the error profile (insertion $i$, deletion $d$, and substitution $s$ probabilities) is estimated using the validation set, following the approach in \cite{srinivasavaradhanTrellisBMACoded2024}. Second, a base model is pretrained over synthetic data generated according to the estimated error profile. This synthetic pretraining strategy is similar to \cite{bar-levScalableRobustDNAbased2025}. Third, the pretrained model is fine-tuned on the validation set, and the final \gls{npd} parameters are used for polar code design via Monte Carlo simulation.

\begin{figure}[!b]
	\centering
	\begin{subfigure}{0.47\textwidth}
		\centering
		\begin{tikzpicture}
\definecolor{chocolate2267451}{RGB}{226,74,51}
\definecolor{dimgray85}{RGB}{85,85,85}
\definecolor{gainsboro229}{RGB}{229,229,229}
\definecolor{darkblue}{RGB}{0, 0, 139}
\definecolor{darkred}{RGB}{139, 0, 0}
\definecolor{darkgreen}{RGB}{0, 100, 0}
\definecolor{darkgray}{RGB}{169, 169, 169}
\definecolor{midnightblue}{RGB}{25, 25, 112}
\definecolor{darkslategray}{RGB}{47, 79, 79}
\definecolor{darkolivegreen}{RGB}{85, 107, 47}
\definecolor{darkviolet}{RGB}{148, 0, 211}
\definecolor{darkorange}{RGB}{255, 140, 0}
\definecolor{darkmagenta}{RGB}{139, 0, 139}
	\begin{axis}[
		width=2.94in,
		height=2.76in,
		ylabel=\textcolor{dimgray85}{BER},
		xlabel=\textcolor{dimgray85}{Sorted Channel Index},
	    grid=major,
		grid style={line width=.2pt, draw=gray!50},
		major grid style={line width=.3pt, draw=gray!50},
		minor tick num=1,
		    xtick={0,128,220,256},
		    ytick={0,0.25,0.5},
		]
		\addplot[
	    only marks,
		mark=*,
		mark size=1.5pt,
		color=darkblue,
		] table[x=X, y=Y, col sep=comma] {./figures/plots/sorted_reliabilities_train_n0-1_best.csv};
		
		\addplot[
		thick,
		color=red,
		] {0} node[pos=0, anchor=south west] {};
		\draw[darkviolet, thick, dashed] (axis cs:220,0) -- (axis cs:220,\pgfkeysvalueof{/pgfplots/ymax});
		
	\end{axis}
\end{tikzpicture}
		\caption{Polarization for punctured real DNA data.}
		\label{fig:dna-nanopore-design-n0-1}
	\end{subfigure}
	\hfill
	\begin{subfigure}{0.47\textwidth}
	\centering
	    \begin{tikzpicture}
	\definecolor{chocolate2267451}{RGB}{226,74,51}
	\definecolor{dimgray85}{RGB}{85,85,85}
	\definecolor{gainsboro229}{RGB}{229,229,229}
	\definecolor{darkblue}{RGB}{0, 0, 139}
	\definecolor{darkred}{RGB}{139, 0, 0}
	\definecolor{darkgreen}{RGB}{0, 100, 0}
	\definecolor{darkgray}{RGB}{169, 169, 169}
	\definecolor{midnightblue}{RGB}{25, 25, 112}
	\definecolor{darkslategray}{RGB}{47, 79, 79}
	\definecolor{darkolivegreen}{RGB}{85, 107, 47}
	\definecolor{darkviolet}{RGB}{148, 0, 211}
	\definecolor{darkorange}{RGB}{255, 140, 0}
	\definecolor{darkmagenta}{RGB}{139, 0, 139}
	\begin{axis}[
		width=2.94in,
		height=2.76in,
		ylabel=\textcolor{dimgray85}{BER},
		xlabel=\textcolor{dimgray85}{Code Rate (bits/base)},
		grid=major,
    log ticks with fixed point,
        ytick={0.005,0.01, 0.05, 0.1},
		xmax=2,
		ymode=log,		
		log basis y={10}, 
		legend pos=north west,
		]
\addplot[
thick,
color=darkgreen,
mark=*,
] coordinates {
	(1.5, 0.0055)
	(1.6, 0.00753)
	(1.7, 0.0094)
	(1.8, 0.0149)
	(1.89, 0.0225)
	(2, 0.0626)
	
};
\addlegendentry{$L=1$}
\addplot[
thick,
color=midnightblue,
mark=*,
] coordinates {
    (1.5, 0.0034)
    (1.6, 0.0051)
	(1.7, 0.0064)
	(1.8, 0.012)
	(1.89, 0.019)
	(2, 0.068)

};
\addlegendentry{$L=8$}

		\end{axis}
	\end{tikzpicture}
	\caption{Decoding \glspl{ber} for \gls{sc} and \gls{scl} decoding.}
	\label{fig:dna-nanopore-trace-dist}
	\end{subfigure}
	\caption{Performance when the number of traces follows $\operatorname{Poiss}(18)$.}
	\label{fig:dna-nanopore-1}
\end{figure}
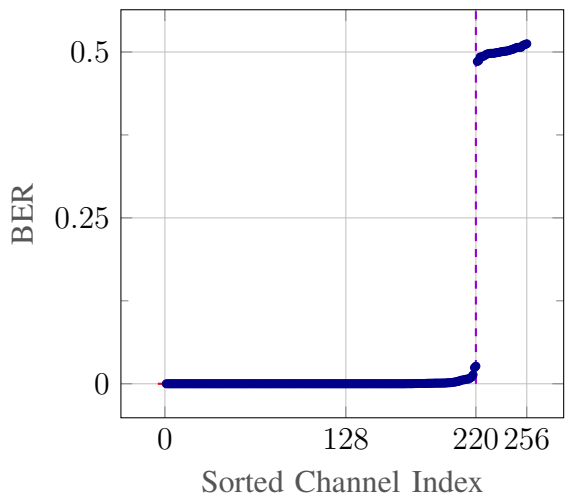
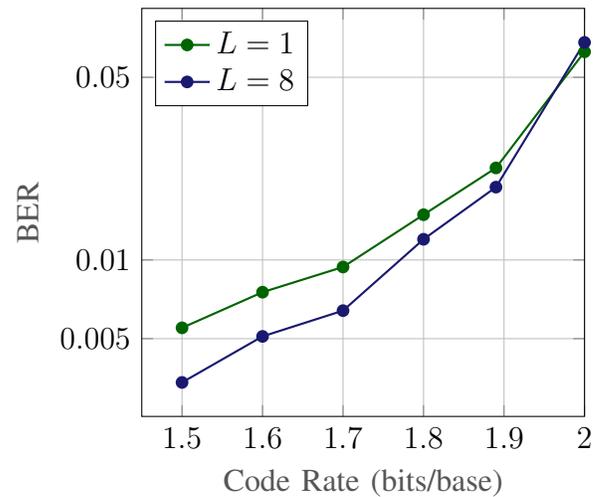

\par A subtlety arises when applying the polar coding scheme to the real Nanopore data. In the dataset, each sample consists of $(\mathbf{x}, \mathbf{y}^{1:K})$, where $\mathbf{x}$ is the transmitted sequence and $\mathbf{y}^{1:K}$ are the observed noisy reads. Therefore, the values of the information and frozen bits were determined independently of our coding design. Nevertheless, since $\mathbf{x}$ was originally sampled uniformly and \gls{iid}, we can reconstruct a polar code structure as follows. We first map $\mathbf{x}$ to its binary representation and pad it with $36$ additional uniformly random bits to create a codeword of length $256$ bits. Then, we apply the inverse Ar\i{}kan transform to obtain a sequence $\mathbf{u}$. 

\par This reconstruction ensures that $\mathbf{u}$ remains uniformly \gls{iid}. The subset of $\mathbf{u}$ corresponding to the designed information set is treated as the information bits, while the complement subset (the frozen set) contains the frozen bits. However, unlike in standard polar coding, the frozen bits are random (not necessarily zero). To resolve this, we assume that the encoder and decoder share randomness (e.g., via a common seed) to consistently regenerate the frozen bits during decoding. Therefore, without loss of generality, the decoder is assumed to know the values of the frozen bits, even though they are not all zeros.

\begin{figure}[!b]
	\centering
	\begin{subfigure}{0.45\textwidth}
		\centering
		\begin{tikzpicture}

	\begin{axis}[
		width=2.94in,
		height=2.76in,
		ylabel=\textcolor{dimgray85}{BER},
		xlabel=\textcolor{dimgray85}{Number of traces},
		grid=major,
	    log ticks with fixed point,
        ytick={0.0001,0.0005, 0.001, 0.005,0.01, 0.05, 0.1, 0.25, 0.5},
		xtick={1,2,3,4,5,6,7,8,9,10},
		ymode=log,		
		log basis y={10}, 
		legend style={
			at={(0.5,1.05)},
			anchor=south,
			font=\small,
			cells={align=left},
			legend columns=3
		},
		]
\addplot[
thick,
color=darkblue,
mark=*, mark size=1.5,
] coordinates {
	(1, 0.484486816)
	(2, 0.470015908)
	(3, 0.424739544)
	(4, 0.361684999)
	(5, 0.305880908)
	(6, 0.250961817)
	(7, 0.207204545)
	(8, 0.178471363)
	(9, 0.141383181)
	(10,0.124162272)
};
 coordinates {
	(1, 0.484486816)
	(2, 0.470015908)
	(3, 0.424739544)
	(4, 0.361684999)
	(5, 0.305880908)
	(6, 0.250961817)
	(7, 0.207204545)
	(8, 0.178471363)
	(9, 0.141383181)
	(10,0.124162272)
};
\addlegendentry{$R=2.0$}
\addplot[
thick,
color=darkred,
mark=*, mark size=1.5,
] coordinates {
	(1,  0.46931)
	(2,  0.39333)
	(3,  0.28454)
	(4,  0.19226)
	(5,  0.12757)
	(6,  0.08776)
	(7,  0.06417)
	(8,  0.04360)
	(9,  0.03438)
	(10, 0.02557)
};
\addlegendentry{$R=1.9$}
\addplot[
thick,
color=darkgreen,
mark=*, mark size=1.5,
] coordinates {
	(1,  0.454147474)
	(2,  0.336751515)
	(3,  0.201860606)
	(4,  0.11375101)
	(5,  0.063215152)
	(6,  0.036037374)
	(7,  0.024116667)
	(8,  0.015410606)
	(9,  0.010421717)
	(10, 0.008307071)
};
\addlegendentry{$R=1.8$}
\addplot[
thick,
color=darkgray,
mark=*, mark size=1.5,
] coordinates {
	(1,  0.442801069)
	(2,  0.258832085)
	(3,  0.137102138)
	(4,  0.073520855)
	(5,  0.031690909)
	(6,  0.02097647)
	(7,  0.01441123)
	(8,  0.007125134)
	(9,  0.005329946)
	(10, 0.003562567)
};
\addlegendentry{$R=1.7$}

\addplot[
thick,
color=midnightblue,
mark=*, mark size=1.5,
] coordinates {
	(1,  0.382496023)
	(2,  0.202592046)
	(3,  0.087501137)
	(4,  0.038736364)
	(5,  0.018091477)
	(6,  0.011988636)
	(7,  0.004915909)
	(8,  0.002838068)
	(9,  0.002560795)
	(10, 0.001613068)
};
\addlegendentry{$R=1.6$}

\addplot[
thick,
color=darkslategray,
mark=*, mark size=1.5,
] coordinates {
	(1,  0.353618182)
	(2,  0.147026667)
	(3,  0.055612121)
	(4,  0.023360606)
	(5,  0.009283636)
	(6,  0.005202424)
	(7,  0.002360606)
	(8,  0.001258788)
	(9,  0.001270909)
	(10, 0.000668485)
};
\addlegendentry{$R=1.5$}

\addplot[
thick,
color=darkolivegreen,
mark=*, mark size=1.5,
] coordinates {
	(1,  0.330405197)
	(2,  0.11135065)
	(3,  0.035092857)
	(4,  0.012281818)
	(5,  0.005259091)
	(6,  0.002344156)
	(7,  0.001378571)
	(8,  0.000751948)
	(9,  0.000428571)
	(10, 0.000264935)
};
\addlegendentry{$R=1.4$}

		\end{axis}
	\end{tikzpicture}
		\caption{Decoding \glspl{ber} for \gls{sc} and \gls{scl} decoding. Number of traces is fixed.}
		\label{fig:dna-nanopore-4}
	\end{subfigure}
	\hfill
	\begin{subfigure}{0.45\textwidth}
	\centering
	    \begin{tikzpicture}
	\definecolor{chocolate2267451}{RGB}{226,74,51}
	\definecolor{dimgray85}{RGB}{85,85,85}
	\definecolor{gainsboro229}{RGB}{229,229,229}
	\definecolor{darkblue}{RGB}{0, 0, 139}
	\definecolor{darkred}{RGB}{139, 0, 0}
	\definecolor{darkgreen}{RGB}{0, 100, 0}
	\definecolor{darkgray}{RGB}{169, 169, 169}
	\definecolor{midnightblue}{RGB}{25, 25, 112}
	\definecolor{darkslategray}{RGB}{47, 79, 79}
	\definecolor{darkolivegreen}{RGB}{85, 107, 47}
	\definecolor{darkviolet}{RGB}{148, 0, 211}
	\definecolor{darkorange}{RGB}{255, 140, 0}
	\definecolor{darkmagenta}{RGB}{139, 0, 139}
    \begin{axis}[
        width=3.04in,
        height=2.76in,
        ybar=4pt,
        bar width=15pt,
        ylabel=\textcolor{dimgray85}{FER},
        ymin=0,
		yticklabels={},
        xtick=data,
        xticklabel style={rotate=45, anchor=east, font=\small},
        xticklabels={NPD, DNAformer, Iterative, BMA Lookahead, Divider BMA, VS},
        nodes near coords={
            \pgfmathprintnumber[fixed,precision=2]{\pgfplotspointmeta}\% 
        },
		nodes near coords align={vertical},
		nodes near coords style={text=dimgray85, font=\footnotesize},
		every axis/.append style={
            label style={text=dimgray85},
            tick label style={text=dimgray85}},
        ]
        \addplot[
            thick,
            color=darkgreen,
            fill=darkgreen,
            ] coordinates {
                (1, 14.94)
                (2, 14.58)
                (3, 16.72)
                (4, 31.44)
                (5, 91.36)
                (6, 94.35)
            };
    \end{axis}
	\end{tikzpicture}
	\caption{Comparison of the \gls{npd} with other methods when no coding is applied.}
	\label{fig:dna-nanopore-5}
	\end{subfigure}
	\caption{Experimental results on real data for the case of a single DNA strand is decoded.}
\end{figure}
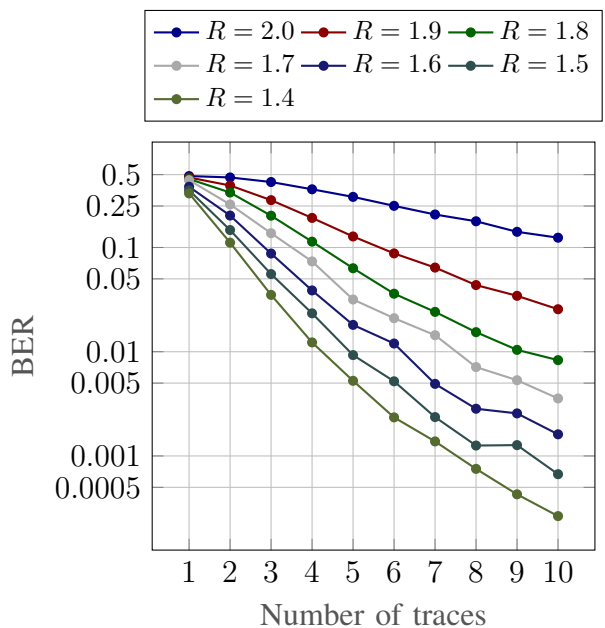
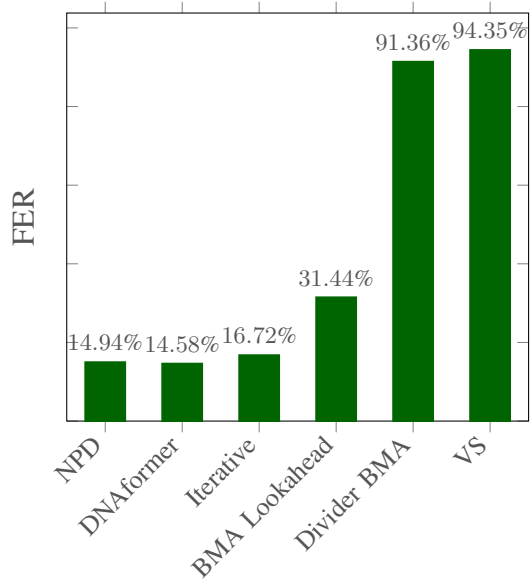
\par For pretraining, we adopt the error profile estimated as $(i,d,s) = (0.017, 0.02, 0.022)$, matching the analysis in \cite{srinivasavaradhanTrellisBMACoded2024}, and set the Poisson sampling parameter $\lambda = 5$. Each fine-tuning iteration on the real validation data proceeds as follows: 
\begin{enumerate}
    \item A batch of $b$ samples $(\mathbf{x}, \mathbf{y}^{1:K})$ is drawn.
    \item For each sample, an independent number of traces $k_b \sim \operatorname{Poiss}(\lambda)$ is sampled.
    \item The minimum of $K$ and $k_b$ traces are retained per sample.
    \item Each $\mathbf{x}$ is converted into binary and padded with $36$ random bits.
    \item The embeddings $\mathbf{e}$ are computed via Equation~\eqref{eqn:multiple_emb}.
    \item The loss is computed according to Equation~\eqref{eqn:npd_loss}.
\end{enumerate}
\par After training, the information set for the polar code is designed by estimating $\mathbb{P}(U_i \neq \hat{u}_i | U^{i-1}=u^{i-1}, \mathbf{Y}=\mathbf{y}^{1:K})$ for each bit and selecting the indices with cumulative error below $0.1$.

\subsubsection{SC Decoding Performance of the NPD for a Single Input Strand}
\par Figure~\ref{fig:dna-nanopore-design-n0-1} shows the polarization behavior for punctured polar codes applied to real DNA data. As expected, about $\frac{36}{256} \approx 0.14$ of the bit-channels are completely noisy due to the random bit padding.

\par Figure~\ref{fig:dna-nanopore-trace-dist} presents the \glspl{ber} achieved by \gls{sc} and \gls{scl} decoding (with list size $8$) at different code rates. Figure~\ref{fig:dna-nanopore-5} compares the decoding performance of our \gls{npd} with other methods, including DNAformer \cite{bar-levScalableRobustDNAbased2025}, the iterative algorithm from \cite{sabaryReconstructionAlgorithmsDNAstorage2024}, and majority-alignment approaches from \cite{sabaryReconstructionAlgorithmsDNAstorage2024,gopalanTraceReconstructionNoisy2017,viswanathanImprovedStringReconstruction2008}, all evaluated under a code rate of $2$ (uncoded). 

\par Although our method slightly underperforms DNAformer in terms of decoding error, it does so with only $3$M parameters, compared to $100$M parameters used by DNAformer, demonstrating the efficiency of the proposed approach.

\subsubsection{SC Decoding Information Encoded on Multiple Input Strands}
\par We now extend the previous single-strand experiment to the case where information is encoded across multiple DNA strands, following the approach in Section~\ref{sec:experiments_synthetic}. This design mitigates variability in trace numbers per strand.

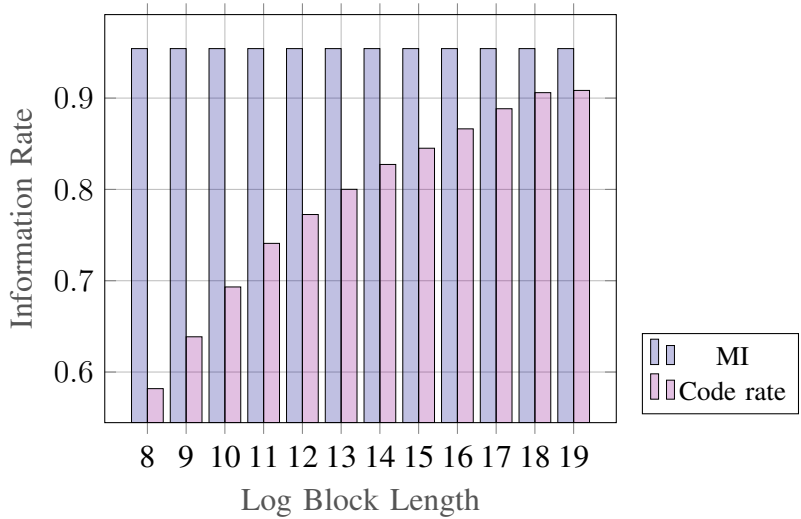
\begin{figure}[t!]
	\centering
	    \begin{tikzpicture}
	\definecolor{chocolate2267451}{RGB}{226,74,51}
	\definecolor{dimgray85}{RGB}{85,85,85}
	\definecolor{gainsboro229}{RGB}{229,229,229}
	\definecolor{darkblue}{RGB}{0, 0, 139}
	\definecolor{darkred}{RGB}{139, 0, 0}
	\definecolor{darkgreen}{RGB}{0, 100, 0}
	\definecolor{darkgray}{RGB}{169, 169, 169}
	\definecolor{midnightblue}{RGB}{25, 25, 112}
	\definecolor{darkslategray}{RGB}{47, 79, 79}
	\definecolor{darkolivegreen}{RGB}{85, 107, 47}
	\definecolor{darkviolet}{RGB}{148, 0, 211}
	\definecolor{darkorange}{RGB}{255, 140, 0}
	\definecolor{darkmagenta}{RGB}{139, 0, 139}
	\begin{axis}[
		width=3.3in,
		height=2.76in,
		ybar=0pt,                        
		bar width=6pt,   
		ylabel=\textcolor{dimgray85}{Information Rate},
		xlabel=\textcolor{dimgray85}{Log Block Length},
		xtick={8,9,10,11,12,13,14,15,16,17,18,19}, 
		xticklabels={8,9,10,11,12,13,14,15,16,17,18,19},
		grid=major,
		legend style={at={(1.05,0.12)}, anchor=west, nodes={scale=0.85, transform shape}},
		]

\addplot+[fill=darkblue, fill opacity=0.25, draw=black, line width=0.25pt] coordinates {
	(8,0.9542)
	(9,0.9542)
	(10,0.9542)
	(11,0.9542)
	(12,0.9542)
	(13,0.9542)
	(14,0.9542)
	(15,0.9542)
	(16,0.9542)
	(17,0.9542)
	(18,0.9542)
	(19,0.9542)
	};
\addlegendentry{MI};
\addplot[fill=darkmagenta, fill opacity=0.25, draw=black, line width=0.25pt
] coordinates {
    (8, 0.58182)
	(9, 0.6386)
	(10, 0.69318)
	(11, 0.740909)
	(12, 0.772500)
	(13, 0.8001)
	(14, 0.827327)
	(15, 0.8451)
	(16, 0.866317)
	(17, 0.888352)
	(18, 0.905957)
	(19, 0.908407)
};
\addlegendentry{Code rate};
		\end{axis}
	\end{tikzpicture}
	\caption{Performance of the \gls{npd} for decoding information stored across multiple concatenated DNA strands on real data. The \glspl{ber} is $\sim0.02$ for all block lengths.}
	\label{fig:dna-nanopore-3}
\end{figure}

\par Figure~\ref{fig:dna-nanopore-3} shows the code rate necessary to achieve a \gls{ber} of approximately $0.01$ and compares it to the estimated average \gls{mi} per bit as a function of the number of concatenated strands $N_0$. We stack up to $2048$ strands, yielding block lengths up to $2^{19}$ bits. 

\par Since the test set contains only $7500$ sequences, we use the following sampling method: for each sample, we draw $N_0$ strands, concatenate their binary representations, and generate a codeword of length $256N_0$ bits. For each concatenated set, the number of traces per strand is sampled as $k \sim \operatorname{Poiss}(5)$, and embeddings are computed independently per trace. The numerical results confirm that the observations made on synthetic DNA data generalize well to real Nanopore data.

\subsubsection{SC Decoding Information Encoded on Multiple Input Strands}
\par We now extend the previous single-strand experiment to the case where information is encoded across multiple DNA strands, following the approach in Section~\ref{sec:experiments_synthetic}. This design mitigates variability in trace numbers per strand.

\par Figure~\ref{fig:dna-nanopore-3} shows the code rate necessary to achieve a \gls{ber} of approximately $0.01$ and compares it to the estimated average \gls{mi} per bit as a function of the number of concatenated strands $N_0$. We stack up to $2048$ strands, yielding block lengths up to $2^{19}$ bits. 

\par Since the test set contains only $7500$ sequences, we use the following sampling method: for each sample, we draw $N_0$ strands, concatenate their binary representations, and generate a codeword of length $256N_0$ bits. For each concatenated set, the number of traces per strand is sampled as $k \sim \operatorname{Poiss}(5)$, and embeddings are computed independently per trace. The numerical results confirm that the observations made on synthetic DNA data generalize well to real Nanopore data.

\subsubsection{Performance under Realistic Sequencing Budget Constraints}
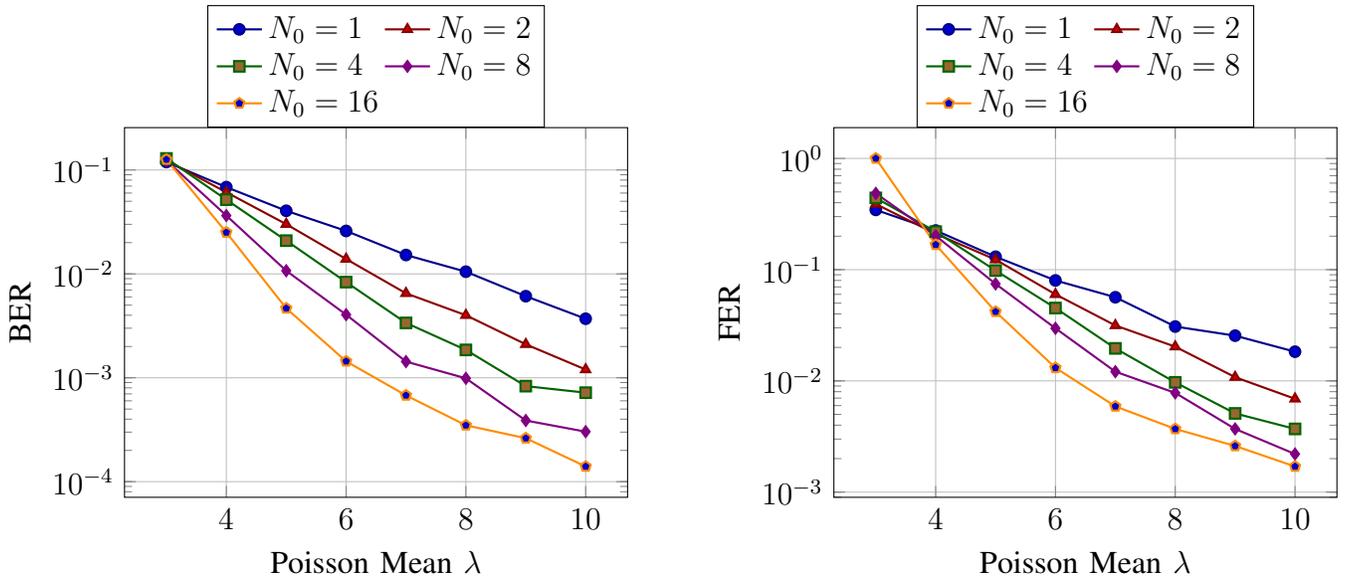
\begin{figure}[t!]
    \centering
    \begin{subfigure}{0.48\textwidth}
        \centering
		\pgfplotstableread[col sep=tab, header=true]{
lambda	fer_n0_1	fer_n0_2	fer_n0_4	fer_n0_8	fer_n0_16
3	0.119300	0.123700	0.129206	0.123404	0.126121
4	0.068400	0.061100	0.051837	0.036429	0.025085
5	0.040500	0.030100	0.020896	0.010724	0.004664
6	0.025900	0.013900	0.008337	0.004051	0.001441
7	0.015200	0.006500	0.003375	0.001432	0.000677
8	0.010500	0.004000	0.001855	0.000988	0.000349
9	0.006100	0.002100	0.000831	0.000388	0.000262
10	0.003700	0.001200	0.000719	0.000303	0.000140
}\bertable

\begin{tikzpicture}
\begin{axis}[
    width=0.95\columnwidth,
    height=6.5cm,
    xlabel={Poisson Mean $\lambda$},
    ylabel={BER},
    ymode=log,
    grid=major,
    legend style={at={(0.5,1.0)}, anchor=south, legend columns=2},
    legend cell align={left}
]

\addplot+[mark=*, thick, color=darkblue]
    table[x=lambda, y=fer_n0_1] {\bertable};
\addlegendentry{$N_0=1$}

\addplot+[mark=triangle*, thick, color=darkred]
    table[x=lambda, y=fer_n0_2] {\bertable};
\addlegendentry{$N_0=2$}

\addplot+[mark=square*, thick, color=darkgreen]
    table[x=lambda, y=fer_n0_4] {\bertable};
\addlegendentry{$N_0=4$}

\addplot+[mark=diamond*, thick, color=violet]
    table[x=lambda, y=fer_n0_8] {\bertable};
\addlegendentry{$N_0=8$}

\addplot+[mark=pentagon*, thick, color=darkorange]
    table[x=lambda, y=fer_n0_16] {\bertable};
\addlegendentry{$N_0=16$}

\end{axis}
\end{tikzpicture}
    \end{subfigure}
    \hfill
    \begin{subfigure}{0.48\textwidth}
        \centering
		\pgfplotstableread[col sep=tab, header=true]{
lambda	fer_n0_1	fer_n0_2	fer_n0_4	fer_n0_8	fer_n0_16
3	0.3447	0.3891	0.4416	0.4854	1.0000
4	0.2250	0.2100	0.2197	0.2018	0.1676
5	0.1307	0.1231	0.0982	0.0745	0.0419
6	0.0800	0.0600	0.0452	0.0297	0.0131
7	0.0565	0.0315	0.0196	0.0121	0.0059
8	0.0308 	0.0203	0.0097	0.0078	0.0037
9	0.0255	0.0108	0.0051	0.0037	0.0026
10	0.0183	0.0069	0.0037	0.0022	0.0017
}\fertable

\begin{tikzpicture}
\begin{axis}[
    width=0.95\columnwidth,
    height=6.5cm,
    xlabel={Poisson Mean $\lambda$},
    ylabel={FER},
    ymode=log,
    grid=major,
    legend style={at={(0.5,1.0)}, anchor=south, legend columns=2},
    legend cell align={left}
]

\addplot+[mark=*, thick, color=darkblue]
    table[x=lambda, y=fer_n0_1] {\fertable};
\addlegendentry{$N_0=1$}

\addplot+[mark=triangle*, thick, color=darkred]
    table[x=lambda, y=fer_n0_2] {\fertable};
\addlegendentry{$N_0=2$}

\addplot+[mark=square*, thick, color=darkgreen]
    table[x=lambda, y=fer_n0_4] {\fertable};
\addlegendentry{$N_0=4$}

\addplot+[mark=diamond*, thick, color=violet]
    table[x=lambda, y=fer_n0_8] {\fertable};
\addlegendentry{$N_0=8$}

\addplot+[mark=pentagon*, thick, color=darkorange]
    table[x=lambda, y=fer_n0_16] {\fertable};
\addlegendentry{$N_0=16$}

\end{axis}
\end{tikzpicture}

    \end{subfigure}
    \caption{Decoding \glspl{ber} and \glspl{fer} for a polar code with rate $1.6$ bits per base on real DNA Nanopore data, shown as a function of the average number of traces per strand ($\lambda$) and the number of strands per codeword ($N_0$).}
    \label{fig:combined-plots}
\end{figure}
\par In practical DNA storage systems, decoding is triggered once the average number of reads per strand reaches a predefined budget, determined by sequencing cost and throughput constraints. This experiment evaluates the decoding performance of the \gls{npd} under such a realistic setting by varying the average number of traces per strand, modeled as the Poisson mean $\lambda$, and the number of concatenated strands $N_0$.

\par For this experiment, we vary $\lambda \in \{3, 4, \dots, 10\}$ and $N_0 \in \{1, 2, 4, 8, 16\}$, where $N_0$ denotes the number of DNA strands concatenated into a single polar codeword of length $256N_0$. For each $(\lambda, N_0)$ pair, decoding is performed using the \gls{npd} with \gls{scl} decoding with list size $8$ on real Nanopore data, and the \glspl{ber} and \gls{fer} are reported. In all experiments we use a code rate of $1.6$ bits per symbol.

\section{Conclusion}\label{sec:conclusion}
\par
This work presents a data-driven decoding approach based on \glspl{npd} for DNA data storage systems, with a particular focus on channels modeled by synchronization errors, such as deletion and \gls{ids} channels. Synchronization errors, common in DNA storage due to synthesis and sequencing noise, disrupt symbol alignment and pose a major challenge for reliable decoding. Existing solutions, like the trellis-based \gls{sc} decoder \cite{talPolarCodesDeletion2021}, enable optimal decoding but suffer from prohibitive $O(N^4\log N)$ complexity.

\par
We propose to use \glspl{npd} to efficiently represent the polar decoding recursion, replacing traditional trellis structures with compact neural network modules. This reduces the decoding complexity to $O(A N \log N)$, where $A$ depends only on the \gls{npd} architecture, achieving significant speedups and enabling practical decoding for long block lengths and list decoding.

\par
While the experiments on deletion and \gls{ids} channels have independent theoretical interest, their primary purpose in this work is to validate \glspl{npd} as viable decoders for realistic DNA data storage systems. The close match between \gls{npd} estimates and known capacity bounds, combined with the successful decoding of synthetic and real DNA data, demonstrates that \glspl{npd} provide a scalable, robust solution for modern DNA storage pipelines.

\par
Overall, this work advances neural decoding architectures for synchronization error channels and opens the door for scalable, data-driven code designs in DNA-based storage. Future directions include extending \glspl{npd} to even more complex DNA error models and integrating outer code optimization into the data-driven training framework.
\clearpage
\appendices
\section{Implementation Details of the Embedding NN}\label{app:implement}
\par This section describes the implementation details of the \gls{npd} in this work. Let $B$ denote the batch size used during training. The hidden size of the \glspl{nn} is denoted by $h$. In all experiments, $d$ and $h$ were determined relative to the block length. Specifically, $d = N/2$ and $h = 2N$.

\subsection{CNN-Based Embedding}\label{app:cnn-embedding}
\par The \gls{cnn} network is one-dimensional and operates along the sequence (block length) axis. Accordingly, each filter has size $f \times d$ with a total of $h$ filters. The filter length is set to $f = \max(N/4, 4)$.

\begin{table}[h]
\small
\centering
\caption{CNN Embedding Architecture}
\label{tab:cnn-embedding}
\begin{tabular}{|c|c|c|c|}
\hline
\textbf{Layer} & \textbf{Input Shape} & \textbf{Output Shape} & \textbf{Parameters} \\
\hline
Padding & $(B, L, 1)$ & $(B, L_\text{max}, 1)$ & -- \\
\hline
Embedding & $(B, L_\text{max}, 1)$ & $(B, L_\text{max}, d)$ & $|\Sigma| d$ \\
\hline
Positional Encoding (added) & $(B, L_\text{max}, d)$ & $(B, L_\text{max}, d)$ & $L_\text{max} d$ \\
\hline
Conv1D (Layer 1) & $(B, L_\text{max}, d)$ & $(B, L_\text{max}, h)$ & $h  f  d$ \\
\hline
Conv1D (Layer 2) & $(B, L_\text{max}, h)$ & $(B, L_\text{max}, h)$ & $h  f  h$ \\
\hline
Conv1D (Layer 3) & $(B, L_\text{max}, h)$ & $(B, L_\text{max}, h)$ & $h  f  h$ \\
\hline
Conv1D (Layer 4) & $(B, L_\text{max}, h)$ & $(B, L_\text{max}, h)$ & $h  f  h$ \\
\hline
Linear Projection & $(B, L_\text{max}, h)$ & $(B, L_\text{max}, d)$ & $h  d + d$ \\
\hline
\end{tabular}
\end{table}
\normalsize
\subsection{Attention-Based Embedding}\label{app:attention-embedding}
\par The attention-based embedding architecture processes the padded and embedded sequence using two attention layers. Each attention layer is followed by a feedforward processing block, referred to as an FC Block. The FC Block consists of a Layer Normalization, a fully connected two-layer MLP with hidden size $h$, and a second Layer Normalization.

\par The input to the first attention layer uses $\mathbf{p}_\mathsf{in}$ as the queries and $\mathbf{\tilde{e}}_\mathsf{pos}$ as both keys and values, where $\mathbf{p}_\mathsf{in} \in \mathbb{R}^{N \times d}$ contains the first $N$ rows of the positional encoding matrix, and $\mathbf{\tilde{e}}_\mathsf{pos} \in \mathbb{R}^{L_\text{max} \times d}$ is the positionally-encoded embedded output sequence. The input to the second attention layer uses the processed representation from the first FC Block as the new queries, while $\mathbf{\tilde{e}}_\mathsf{pos}$ remains the keys and values. Table~\ref{tab:attention-embedding} summarizes the full layer structure, dimensions, and trainable parameters.

\begin{table}[h]
\small
\centering
\caption{Attention Embedding Architecture}
\label{tab:attention-embedding}
\renewcommand{\arraystretch}{1.1}
\begin{tabular}{|c|c|c|c|}
\hline
\textbf{Layer} & \textbf{Input Shape} & \textbf{Output Shape} & \textbf{Parameters} \\
\hline
Padding & $(B, L, 1)$ & $(B, L_\text{max}, 1)$ & -- \\
\hline
Embedding & $(B, L_\text{max}, 1)$ & $(B, L_\text{max}, d)$ & $|\Sigma|  d$ \\
\hline
Positional Encoding & $(B, L_\text{max}, d)$ & $(B, L_\text{max}, d)$ & $L_\text{max}  d$ \\
\hline
Self-Attention (Layer 1) & $(B, N, d)$, $(B, L_\text{max}, d)$ & $(B, N, d)$ & $4d^2$ \\
\hline
FC Block 1 & $(B, N, d)$ & $(B, N, d)$ & $2d + (2dh + h + d) + 2d$ \\
\hline
Cross-Attention (Layer 2) & $(B, N, d)$, $(B, L_\text{max}, d)$ & $(B, N, d)$ & $4d^2$ \\
\hline
FC Block 2 & $(B, N, d)$ & $(B, N, d)$ & $2d + (2dh + h + d) + 2d$ \\
\hline
\end{tabular}
\end{table}
\normalsize

\section{Implementation Details of $F_\theta$, $G_\theta$, and $H_\theta$}

\par In this section, we describe the implementation details of the remaining three neural network components of the \gls{npd}: the check-node network $F_\theta$, the bit-node network $G_\theta$, and the embedding-to-LLR network $H_\theta$.

\subsection{Check-Node Network $F_\theta$}

\par The check-node function $F_\theta$ maps two embeddings into a new embedding. The architecture of $F_\theta$ is illustrated in the following table.

\begin{table}[h]
\small
\centering
\caption{Check-Node $F_\theta$ Architecture}
\label{tab:check-node}
\begin{tabular}{|c|c|c|c|}
\hline
\textbf{Layer} & \textbf{Input Shape} & \textbf{Output Shape} & \textbf{Parameters} \\
\hline
Input Concatenation & $(B, N, d), (B, N, d)$ & $(B, N, 2d)$ & -- \\
\hline
Fully Connected Layer 1 & $(B, N, 2d)$ & $(B, N, h)$ & $2dh + h$ \\
\hline
ReLU Activation & $(B, N, h)$ & $(B, N, h)$ & -- \\
\hline
Fully Connected Layer 2 & $(B, N, h)$ & $(B, N, d)$ & $hd + d$ \\
\hline
\end{tabular}
\end{table}
\normalsize

\subsection{Bit-Node Network $G_\theta$}

\par The bit-node function $G_\theta$ maps two embeddings and the previous bit decision into a new embedding. The previous bit $u \in \{0,1\}$ is first embedded into a $d$-dimensional vector using a learned embedding layer. Thus, the input to $G_\theta$ after concatenation has shape $(B, N, 3d)$. The architecture of $G_\theta$ is illustrated in the following table.

\begin{table}[h]
\small
\centering
\caption{Bit-Node $G_\theta$ Architecture}
\label{tab:bit-node}
\begin{tabular}{|c|c|c|c|}
\hline
\textbf{Layer} & \textbf{Input Shape} & \textbf{Output Shape} & \textbf{Parameters} \\
\hline
Bit Embedding & $(B, N, 1)$ & $(B, N, d)$ & $2d$ \\
\hline
Input Concatenation & $(B, N, d), (B, N, d), (B, N, d)$ & $(B, N, 3d)$ & -- \\
\hline
Fully Connected Layer 1 & $(B, N, 3d)$ & $(B, N, h)$ & $3dh + h$ \\
\hline
ReLU Activation & $(B, N, h)$ & $(B, N, h)$ & -- \\
\hline
Fully Connected Layer 2 & $(B, N, h)$ & $(B, N, d)$ & $hd + d$ \\
\hline
\end{tabular}
\end{table}
\normalsize

\subsection{Embedding-to-LLR Network $H_\theta$}

\par The embedding-to-LLR network $H_\theta$ maps an embedding vector into a scalar log-likelihood ratio (LLR). It serves as the final step before making a hard decision on the bit value. The architecture of $H_\theta$ is illustrated in the following table.

\begin{table}[h]
\small
\centering
\caption{Embedding-to-LLR $H_\theta$ Architecture}
\label{tab:llr-node}
\begin{tabular}{|c|c|c|c|}
\hline
\textbf{Layer} & \textbf{Input Shape} & \textbf{Output Shape} & \textbf{Parameters} \\
\hline
Fully Connected Layer 1 & $(B, N, d)$ & $(B, N, h)$ & $dh + h$ \\
\hline
ReLU Activation & $(B, N, h)$ & $(B, N, h)$ & -- \\
\hline
Fully Connected Layer 2 & $(B, N, h)$ & $(B, N, 1)$ & $h + 1$ \\
\hline
\end{tabular}
\end{table}
\normalsize
\printbibliography

@inproceedings{aharoniCodeRateOptimization2024,
  title = {Code {{Rate Optimization}} via {{Neural Polar Decoders}}},
  booktitle = {2024 {{IEEE International Symposium}} on {{Information Theory}} ({{ISIT}})},
  author = {Aharoni, Ziv and Huleihel, Bashar and Pfister, Henry D. and Permuter, Haim H.},
  year = {2024},
  pages = {2424--2429},
  issn = {2157-8117},
  doi = {10.1109/ISIT57864.2024.10619429}
}

@article{aharoniDatadrivenNeuralPolar2024,
  title = {Data-Driven {{Neural Polar Decoders}} for {{Unknown Channels}} with and without {{Memory}}},
  author = {Aharoni, Ziv and Huleihel, Bashar and Pfister, Henry D and Permuter, Haim H},
  year = {2024},
  journal = {IEEE Transactions on Information Theory},
  publisher = {IEEE}
}

@article{arikanChannelPolarizationMethod2009,
  title = {Channel {{Polarization}}: {{A Method}} for {{Constructing Capacity-achieving Codes}} for {{Symmetric Binary-input Memoryless Channels}}},
  author = {Arikan, E.},
  year = {2009},
  journal = {IEEE Trans. Inf. Theory},
  volume = {55},
  number = {7},
  pages = {3051--3073},
  publisher = {IEEE}
}

@article{bancroftLongtermStorageInformation2001,
  title = {Long-Term Storage of Information in {{DNA}}},
  author = {Bancroft, Carter and Bowler, Timothy and Bloom, Brian and Clelland, Catherine Taylor},
  year = {2001},
  journal = {Science},
  volume = {293},
  number = {5536},
  pages = {1763--1765},
  publisher = {American Association for the Advancement of Science}
}

@article{bar-levScalableRobustDNAbased2025,
  title = {Scalable and Robust {{DNA-based}} Storage via Coding Theory and Deep Learning},
  author = {{Bar-Lev}, Daniella and Orr, Itai and Sabary, Omer and Etzion, Tuvi and Yaakobi, Eitan},
  year = {2025},
  journal = {Nat Mach Intell},
  pages = {1--11},
  publisher = {Nature Publishing Group},
  issn = {2522-5839},
  doi = {10.1038/s42256-025-01003-z}
}

@inproceedings{batu2004reconstructing,
  title = {Reconstructing Strings from Random Traces},
  booktitle = {{{SODA}}},
  author = {Batu, Tugkan and Kannan, Sampath and Khanna, Sanjeev and McGregor, Andrew},
  year = {2004},
  volume = {4},
  pages = {910--918}
}

@inproceedings{castiglioneTrellisBasedLower2015,
  title = {Trellis Based Lower Bounds on Capacities of Channels with Synchronization Errors},
  booktitle = {2015 {{IEEE Information Theory Workshop-Fall}} ({{ITW}})},
  author = {Castiglione, Jason and Kavcic, Aleksandar},
  year = {2015},
  pages = {24--28},
  publisher = {IEEE}
}

@article{churchNextgenerationDigitalInformation2012,
  title = {Next-Generation Digital Information Storage in {{DNA}}},
  author = {Church, George M and Gao, Yuan and Kosuri, Sriram},
  year = {2012},
  journal = {Science},
  volume = {337},
  number = {6102},
  pages = {1628--1628},
  publisher = {American Association for the Advancement of Science}
}

@article{daveyReliableCommunicationChannels2001,
  title = {Reliable Communication over Channels with Insertions, Deletions, and Substitutions},
  author = {Davey, Matthew C and MacKay, David JC},
  year = {2001},
  journal = {IEEE Transactions on Information Theory},
  volume = {47},
  number = {2},
  pages = {687--698},
  publisher = {IEEE}
}

@article{dobrushinShannonsTheoremsChannels1967,
  title = {Shannon's Theorems for Channels with Synchronization Errors},
  author = {Dobrushin, Roland L'vovich},
  year = {1967},
  journal = {Problemy Peredachi Informatsii},
  volume = {3},
  number = {4},
  pages = {18--36},
  publisher = {{Russian Academy of Sciences, Branch of Informatics, Computer Equipment and {\dots}}}
}

@misc{erlichDNAFountainEnables2017,
  title = {{{DNA Fountain}} Enables a Robust and Efficient Storage Architecture},
  author = {Erlich, Yaniv and Zielinski, Dina},
  year = {2017},
  doi = {10.1126/science.aaj2038},
  howpublished = {https://www.science.org/doi/10.1126/science.aaj2038}
}

@article{fertonaniBoundsCapacityChannels2011,
  title = {Bounds on the {{Capacity}} of {{Channels}} with {{Insertions}}, {{Deletions}} and {{Substitutions}}},
  author = {Fertonani, Dario and Duman, Tolga M. and Erden, M. Fatih},
  year = {2011},
  journal = {IEEE Transactions on Communications},
  volume = {59},
  number = {1},
  pages = {2--6},
  issn = {1558-0857},
  doi = {10.1109/TCOMM.2010.110310.090039}
}

@book{gallagerSequentialDecodingBinary1961,
  title = {Sequential Decoding for Binary Channels with Noise and Synchronization Errors},
  author = {Gallager, Robert G},
  year = {1961},
  publisher = {British Library, Reports \& Microfilms}
}

@misc{gehringConvolutionalSequenceSequence2017,
  title = {Convolutional {{Sequence}} to {{Sequence Learning}}},
  author = {Gehring, Jonas and Auli, Michael and Grangier, David and Yarats, Denis and Dauphin, Yann N.},
  year = {2017},
  number = {arXiv:1705.03122},
  eprint = {1705.03122},
  primaryclass = {cs},
  publisher = {arXiv},
  doi = {10.48550/arXiv.1705.03122},
  archiveprefix = {arXiv}
}

@patent{gopalanTraceReconstructionNoisy2017,
  title = {Trace Reconstruction from Noisy Polynucleotide Sequencer Reads},
  author = {GOPALAN, Parikshit S. and Yekhanin, Sergey and ANG, Siena Dumas and Jojic, Nebojsa and Racz, Miklos and Strauss, Karin and Ceze, Luis},
  year = {2017},
  number = {WO2017189469A1},
  assignee = {Microsoft Technology Licensing, Llc},
  nationality = {WO}
}

@article{grassRobustChemicalPreservation2015,
  title = {Robust {{Chemical Preservation}} of {{Digital Information}} on {{DNA}} in {{Silica}} with {{Error-Correcting Codes}}},
  author = {Grass, Robert N. and Heckel, Reinhard and Puddu, Michela and Paunescu, Daniela and Stark, Wendelin J.},
  year = {2015},
  journal = {Angewandte Chemie International Edition},
  volume = {54},
  number = {8},
  pages = {2552--2555},
  issn = {1521-3773},
  doi = {10.1002/anie.201411378}
}

@misc{heckelCharacterizationDNAData2018,
  title = {A {{Characterization}} of the {{DNA Data Storage Channel}}},
  author = {Heckel, Reinhard and Mikutis, Gediminas and Grass, Robert N.},
  year = {2018},
  number = {arXiv:1803.03322},
  eprint = {1803.03322},
  primaryclass = {cs},
  publisher = {arXiv},
  doi = {10.48550/arXiv.1803.03322},
  archiveprefix = {arXiv}
}

@inproceedings{heckelFundamentalLimitsDNA2017a,
  title = {Fundamental Limits of {{DNA}} Storage Systems},
  booktitle = {2017 {{IEEE International Symposium}} on {{Information Theory}} ({{ISIT}})},
  author = {Heckel, Reinhard and Shomorony, Ilan and Ramchandran, Kannan and Tse, David N. C.},
  year = {2017},
  pages = {3130--3134},
  issn = {2157-8117},
  doi = {10.1109/ISIT.2017.8007106}
}

@inproceedings{lenzAchievingCapacityDNA2020,
  title = {Achieving the {{Capacity}} of the {{DNA Storage Channel}}},
  booktitle = {{{ICASSP}} 2020 - 2020 {{IEEE International Conference}} on {{Acoustics}}, {{Speech}} and {{Signal Processing}} ({{ICASSP}})},
  author = {Lenz, Andreas and Siegel, Paul H. and {Wachter-Zeh}, Antonia and Yaakohi, Eitan},
  year = {2020},
  pages = {8846--8850},
  issn = {2379-190X},
  doi = {10.1109/ICASSP40776.2020.9053049}
}

@article{lenzCodingSetsDNA2020,
  title = {Coding {{Over Sets}} for {{DNA Storage}}},
  author = {Lenz, Andreas and Siegel, Paul H. and {Wachter-Zeh}, Antonia and Yaakobi, Eitan},
  year = {2020},
  journal = {IEEE Transactions on Information Theory},
  volume = {66},
  number = {4},
  pages = {2331--2351},
  issn = {1557-9654},
  doi = {10.1109/TIT.2019.2961265}
}

@article{leproustSynthesisHighqualityLibraries2010,
  title = {Synthesis of High-Quality Libraries of Long (150mer) Oligonucleotides by a Novel Depurination Controlled Process},
  author = {LeProust, Emily M. and Peck, Bill J. and Spirin, Konstantin and McCuen, Heather Brummel and Moore, Bridget and Namsaraev, Eugeni and Caruthers, Marvin H.},
  year = {2010},
  journal = {Nucleic Acids Research},
  volume = {38},
  number = {8},
  pages = {2522--2540},
  issn = {0305-1048, 1362-4962},
  doi = {10.1093/nar/gkq163}
}

@article{levenshteinEfficientReconstructionSequences2001,
  title = {Efficient Reconstruction of Sequences},
  author = {Levenshtein, V.I.},
  year = {2001},
  journal = {IEEE Transactions on Information Theory},
  volume = {47},
  number = {1},
  pages = {2--22},
  issn = {1557-9654},
  doi = {10.1109/18.904499}
}

@article{meiserSyntheticDNAApplications2022,
  title = {Synthetic {{DNA}} Applications in Information Technology},
  author = {Meiser, Linda C. and Nguyen, Bichlien H. and Chen, Yuan-Jyue and Nivala, Jeff and Strauss, Karin and Ceze, Luis and Grass, Robert N.},
  year = {2022},
  journal = {Nat Commun},
  volume = {13},
  number = {1},
  pages = {352},
  publisher = {Nature Publishing Group},
  issn = {2041-1723},
  doi = {10.1038/s41467-021-27846-9}
}

@misc{mnihAsynchronousMethodsDeep2016,
  title = {Asynchronous {{Methods}} for {{Deep Reinforcement Learning}}},
  author = {Mnih, Volodymyr and Badia, Adri{\`a} Puigdom{\`e}nech and Mirza, Mehdi and Graves, Alex and Lillicrap, Timothy P. and Harley, Tim and Silver, David and Kavukcuoglu, Koray},
  year = {2016},
  number = {arXiv:1602.01783},
  eprint = {1602.01783},
  primaryclass = {cs},
  publisher = {arXiv},
  doi = {10.48550/arXiv.1602.01783},
  archiveprefix = {arXiv}
}

@article{organickRandomAccessLargescale2018,
  title = {Random Access in Large-Scale {{DNA}} Data Storage},
  author = {Organick, Lee and Ang, Siena Dumas and Chen, Yuan-Jyue and Lopez, Randolph and Yekhanin, Sergey and Makarychev, Konstantin and Racz, Miklos Z and Kamath, Govinda and Gopalan, Parikshit and Nguyen, Bichlien and Takahashi, Christopher N and Newman, Sharon and Parker, Hsing-Yeh and Rashtchian, Cyrus and Stewart, Kendall and Gupta, Gagan and Carlson, Robert and Mulligan, John and Carmean, Douglas and Seelig, Georg and Ceze, Luis and Strauss, Karin},
  year = {2018},
  journal = {Nat Biotechnol},
  volume = {36},
  number = {3},
  pages = {242--248},
  issn = {1087-0156, 1546-1696},
  doi = {10.1038/nbt.4079}
}

@article{perniceMutualInformationUpper2024,
  title = {Mutual Information Upper Bounds for Uniform Inputs through the Deletion Channel},
  author = {Pernice, Francisco and Isik, Berivan and Weissman, Tsachy},
  year = {2024},
  journal = {IEEE Transactions on Information Theory},
  publisher = {IEEE}
}

@article{rahmatiUpperBoundsCapacity2015,
  title = {Upper {{Bounds}} on the {{Capacity}} of {{Deletion Channels Using Channel Fragmentation}}},
  author = {Rahmati, Mojtaba and Duman, Tolga M.},
  year = {2015},
  journal = {IEEE Transactions on Information Theory},
  volume = {61},
  number = {1},
  pages = {146--156},
  issn = {1557-9654},
  doi = {10.1109/TIT.2014.2368553}
}

@article{sabaryReconstructionAlgorithmsDNAstorage2024,
  title = {Reconstruction Algorithms for {{DNA-storage}} Systems},
  author = {Sabary, Omer and Yucovich, Alexander and Shapira, Guy and Yaakobi, Eitan},
  year = {2024},
  journal = {Sci Rep},
  volume = {14},
  number = {1},
  pages = {1951},
  publisher = {Nature Publishing Group},
  issn = {2045-2322},
  doi = {10.1038/s41598-024-51730-3}
}

@article{sabarySurveyDecadeCoding2024,
  title = {Survey for a {{Decade}} of {{Coding}} for {{DNA Storage}}},
  author = {Sabary, Omer and Kiah, Han Mao and Siegel, Paul H. and Yaakobi, Eitan},
  year = {2024},
  journal = {IEEE Transactions on Molecular, Biological, and Multi-Scale Communications},
  volume = {10},
  number = {2},
  pages = {253--271},
  issn = {2332-7804},
  doi = {10.1109/TMBMC.2024.3403488}
}

@misc{srinivasavaradhanTrellisBMACoded2024,
  title = {Trellis {{BMA}}: {{Coded Trace Reconstruction}} on {{IDS Channels}} for {{DNA Storage}}},
  shorttitle = {Trellis {{BMA}}},
  author = {Srinivasavaradhan, Sundara Rajan and Gopi, Sivakanth and Pfister, Henry D. and Yekhanin, Sergey},
  year = {2024},
  number = {arXiv:2107.06440},
  eprint = {2107.06440},
  primaryclass = {cs},
  publisher = {arXiv},
  doi = {10.48550/arXiv.2107.06440},
  archiveprefix = {arXiv}
}

@article{talPolarCodesDeletion2021,
  title = {Polar {{Codes}} for the {{Deletion Channel}}: {{Weak}} and {{Strong Polarization}}},
  author = {Tal, I. and Pfister, H. D. and Fazeli, A. and Vardy, A.},
  year = {2021},
  journal = {IEEE Trans. Inf. Theory},
  volume = {68},
  number = {4},
  pages = {2239--2265},
  publisher = {IEEE}
}

@article{varshamovCodesWhichCorrect1965a,
  title = {Codes Which Correct Single Asymmetric Errors},
  author = {Varshamov, R. R. and Tenengolts, G. M.},
  year = {1965},
  journal = {Automation and Remote Control},
  volume = {26},
  number = {2},
  pages = {286--290}
}

@inproceedings{vaswaniAttentionAllYou2017,
  title = {Attention Is {{All}} You {{Need}}},
  booktitle = {Advances in {{Neural Information Processing Systems}}},
  author = {Vaswani, Ashish and Shazeer, Noam and Parmar, Niki and Uszkoreit, Jakob and Jones, Llion and Gomez, Aidan N and ukasz Kaiser, {\L} and Polosukhin, Illia},
  editor = {Guyon, I. and Luxburg, U. Von and Bengio, S. and Wallach, H. and Fergus, R. and Vishwanathan, S. and Garnett, R.},
  year = {2017},
  volume = {30},
  publisher = {Curran Associates, Inc.}
}

@inproceedings{viswanathanImprovedStringReconstruction2008,
  title = {Improved String Reconstruction over Insertion-Deletion Channels},
  booktitle = {Proceedings of the Nineteenth Annual {{ACM-SIAM}} Symposium on {{Discrete}} Algorithms},
  author = {Viswanathan, Krishnamurthy and Swaminathan, Ram},
  year = {2008},
  series = {{{SODA}} '08},
  pages = {399--408},
  publisher = {{Society for Industrial and Applied Mathematics}},
  address = {USA}
}

@inproceedings{wangConstructionPolarCodes2015,
  title = {Construction of {{Polar Codes}} for {{Channels}} with {{Memory}}},
  booktitle = {2015 {{IEEE Information Theory Workshop-Fall}} ({{ITW}})},
  author = {Wang, R. and Honda, J. and Yamamoto, H. and Liu, R. and Hou, Y.},
  year = {2015},
  pages = {187--191},
  publisher = {IEEE}
}

\end{document}